%% file: DQHO_main.tex
\documentclass[reprint,
nofootinbib,
 amsmath,amssymb,
 aps,
pra,
floatfix,
]{revtex4-2}
\usepackage[dvipsnames,svgnames]{xcolor}
\usepackage[colorlinks=true,citecolor=blue,linkcolor=blue]{hyperref}
\usepackage{graphicx}% Include figure files
\usepackage{dcolumn}% Align table columns on decimal point
\usepackage{bm}% bold math
\usepackage{amsmath}
\usepackage{cleveref}
\usepackage{amssymb}
\usepackage[svgnames]{xcolor}
\usepackage{mathtools}
\usepackage{braket}
\usepackage{physics}
\usepackage{subfiles}
\usepackage{float}
\def\Xint#1{\mathchoice
   {\XXint\displaystyle\textstyle{#1}}%
   {\XXint\textstyle\scriptstyle{#1}}%
   {\XXint\scriptstyle\scriptscriptstyle{#1}}%
   {\XXint\scriptscriptstyle\scriptscriptstyle{#1}}%
   \!\int}
\def\XXint#1#2#3{{\setbox0=\hbox{$#1{#2#3}{\int}$}
     \vcenter{\hbox{$#2#3$}}\kern-.5\wd0}}
\def\ddashint{\Xint=}
\def\dashint{\Xint-}
\def\floor#1{\left\lfloor #1 \right \rfloor}

\newcommand{\emb}{\text{em}}
\newcommand{\ret}{\text{ret}}

\newcommand{\cl}{{\rm cl}}
\newcommand{\q}{{\rm q}}
\newcommand{\Var}{\text{Var}}
\newcommand{\inv}{^{\raisebox{.2ex}{$\scriptscriptstyle-1$}}}
\newcommand{\sgn}{\mathrm{sgn}}
\renewcommand{\adv}{\text{adv}}
\newcommand{\fp}{\text{f.p.}}
\newcommand{\sinc}{\text{sinc}}
\begin{document}

%Nonequilibrium Green’s Function Theory for Multiscale Open Quantum Systems
\title{Hadamard regularization of open quantum systems coupled to unstructured environments in the Schwinger-Keldysh formalism}% Force line breaks with \\

\author{Jakob Dolgner}
 \email{j.dolgner@fkf.mpg.de}%Lines break automatically or can be forced with \\
%\author{Dirk Manske}%
 %\email{d.manske@fkf.mpg.de}
\affiliation{%
 Quantum Many-Body Theory Department, Max Planck Institute for Solid State Research
}%
\date{\today}% It is always \today, today,

\begin{abstract}
The theory of open quantum systems addresses how coupling to external degrees of freedom modifies observables and quantum coherence, a situation central to fundamental condensed-matter research and emerging quantum technologies. Schwinger-Keldysh field theory is a natural framework for both open- and nonequilibrium quantum systems in terms of functional integrals. However, its numerical solution is limited by a cubic scaling with the number of time steps. This is particularly prohibitive for scenarios with widely separated time scales, as is often the case for system and environmental scales. We consider a damped quantum harmonic oscillator as a toy model to study a separation-of-scales ansatz based on Hadamard regularization. A time-stepping algorithm for the Kadanoff-Baym equations on the slow system time-scale is presented that captures both low-temperature non-Markovianity and renormalization effects arising from the much faster environment scale. 
\end{abstract}

\keywords{ohmic damping, damped quantum harmonic oscillator, Caldeira-Leggett model, Keldysh formalism, Hadamard finite-part, Non-Markovian, embedding self-energy, separation-of-scales, Kadanoff-Baym equations}%Use showkeys class option if keyword
\maketitle

%\tableofcontents

\subfile{sections/introduction}

\subfile{sections/Equilibrium}

\subfile{sections/Nonequilibrium}

\subfile{sections/Separation_of_Scales}

\subfile{sections/Implementation}

\subfile{sections/conclusion}

%\twocolumngrid
\subfile{sections/Acknowledgment}

\bibliography{biblio}

\newpage
\appendix

\onecolumngrid

\section{The Caldeira-Leggett environment}\label{sec:Caldeira-Legget}
\subfile{sections/appendices/Caldeira_Leggett_model}

\section{Finite mass renormalization}\label{sec:mass renorm}
\subfile{sections/appendices/finite_wc_mass_renorm}

\section{Fourier transform of the self-energy}\label{sec:Fourier_derivation}
\subfile{sections/appendices/Fourier_trf}

\section{Principal value and finite part identities for ohmic bath memory kernels}\label{sec:HFP and CPV identities}
\subfile{sections/appendices/HFP_identities}

\section{The Hadamard finite part}\label{sec:HFP}
\subfile{sections/appendices/finite_part}

\section{Nascent $\delta^2$}\label{sec:delta_squared_derivation}
\subfile{sections/appendices/nascent_delta2}

\section{Analytic results for the zero- and one-terms, $P$ and $Q$}\label{sec:zero_and_one}
\subfile{sections/appendices/zero_and_one_terms}

%\section{Langreth rules}\label{sec:langreth rules}
%\subfile{sections/appendices/Langreth_rules}

\section{Exponential quadrature method}\label{sec:ETD}
\subfile{sections/appendices/ETD}

\section{Imaginary time as compactified past}\label{sec:compact_past}
\subfile{sections/appendices/t0_shifting}

% \section{Derivative Linearization}\label{sec:derivative_linearization}
% \subfile{sections/appendices/Linearization}

\end{document}

%% file: sections/introduction.tex
\section{\label{sec:Introduction}Introduction}
The study of damped quantum systems attracted significant interest in the early development of quantum mechanics and quantum field theory \cite{weisskopf_berechnung_1930,schwinger_brownian_1961}. Classically ubiquitous phenomena, such as damping and irreversibility, required reconciliation with the fundamental unitarity of quantum mechanics. It is now well-established that the solution lies in the framework of open quantum systems, defined as a system of interest coupled to an environment \cite{senitzky_dissipation_1960,ford_quantum_1988}. While the time evolution of the total bipartite state in $\mathcal{H}_\text{sys}\otimes \mathcal{H}_\text{env}$ remains unitary, the system-environment coupling generates entanglement. Once the environment is traced out, this manifests as irreversible and generally non-Markovian dynamics in the reduced system \cite{breuer_colloquium_2016, de_vega_dynamics_2017}. Theoretical approaches to open quantum systems can be broadly categorized into "state-based methods", which solve for the reduced density matrix, and "observable-based methods", which focus on operator expectation values and Green's functions. The state-based class includes the formally exact Nakajima-Zwanzig projection technique \cite{nakajima_quantum_1958, zwanzig_ensemble_1960} and its various practical implementations. These range from the perturbative Redfield \cite{redfield_theory_1957} and Lindblad \cite{lindblad_generators_1976} master equations to the numerically exact hierarchical equations of motion (HEOM) \cite{tanimura_time_1989} and stochastic Schrödinger equation approaches \cite{plenio_quantum-jump_1998}. The second class encompasses path-integral and field-theoretic techniques, such as the Feynman-Vernon influence functional \cite{feynman_theory_1963}, the Quantum Langevin equation \cite{ford_quantum_1987, ford_quantum_1988}, and the Schwinger-Keldysh formalism \cite{schwinger_brownian_1961,baym_conservation_1961,Keldysh_diagram_1965}. The choice of method depends critically on the problem at hand. State-based methods typically scale unfavorably with the Hilbert space dimension, whereas observable-based approaches often face challenges with strong correlations in interacting systems. Furthermore, access to multi-time correlation functions, which are indispensable for spectroscopy and transport properties, further constrains the set of viable theoretical frameworks. While most state-based approaches rely on the Quantum Regression Theorem (QRT), which is known to fail in general non-Markovian regimes \cite{talkner_failure_1986}, the Schwinger-Keldysh formalism naturally provides access to these quantities. It stands alongside the Quantum Langevin equation (for linear systems) and extended-state methods such as HEOM as one of the few rigorous approaches to exact nonequilibrium correlation functions \cite{tanimura_stochastic_2006}.

In quantum optics and condensed matter physics, open quantum systems are the rule rather than the exception \cite{breuer_theory_2007, weiss_quantum_2021}. Specifically, in the context of nonequilibrium dynamics, the system's openness dictates the time-scales of decoherence and thermalization \cite{gardiner_quantum_2004, zurek_decoherence_2003}. A standard first approximation is to introduce a phenomenological damping by embedding the system in a structureless environment. In the state-based formalism, this is achieved efficiently by generalizing unitary dynamics into a Lindblad master equation. This approach is highly effective provided the inherent Born-Markov approximation is justified. An analogous embedding is possible in the Schwinger-Keldysh formalism through the addition of embedding self-energies\footnote{Historically, Schwinger developed the closed-time-path formalism (1961) to describe the Brownian motion of a quantum particle \cite{schwinger_brownian_1961}, predating Keldysh's (1964) independent application of the same framework to nonequilibrium field theory \cite{Keldysh_diagram_1965}.} \cite{kamenev_field_2011, stefanucci_nonequilibrium_2013}. While formally exact (in practice, a truncation of the Schwinger-Dyson hierarchy is necessary) \cite{weiss_quantum_2021}, a computational bottleneck arises when implementing the featureless environmental limit. Unlike in the state-based formalism, where featureless rates simplify the dynamics, in the Kadanoff-Baym equations (KBE), they lead to divergences. The standard phenomenological self-energies, the wide-band limit for fermions and the Ohmic limit for bosons, take the form:
\begin{equation}\Sigma^\ret(\omega) = - i\gamma \quad \text{and} \quad \Sigma^\ret(\omega) = - i \gamma \omega\,,\end{equation}
respectively. Their unbounded spectra generate singularities, some of which persist even after renormalization via counter-terms \cite{caldeira_influence_1981,caldeira_path_1983}. Consequently, one is forced to regularize the theory by introducing a high-frequency cutoff $\omega_c$, modifying the rates with regulating factors such as \cite{grabert_quantum_1984,grabert_localization_1987,grabert_quantum_1988}
\begin{equation}\theta(\omega_c-\abs{\omega})\,,\quad e^{-\frac{\abs{\omega}}{\omega_c}}\,,\quad \text{or} \quad \frac{\omega_c^2}{\omega^2 + \omega_c^2}\,.\end{equation}
Crucially, to approximate a featureless bath, $\omega_c$ must far exceed the system's characteristic energy scales. This introduces a stiff hierarchy of time scales, forcing the KBE solver to adopt extremely small time steps to resolve $1/\omega_c$, thereby prohibitively increasing the computational cost. 
To circumvent this ensuing stiffness, state-of-the-art numerical schemes employ a sum-over-poles decomposition of the thermal distribution function. This technique allows the wide-band self-energies to be expressed as a sum of exponentials, thereby converting the intractable memory integrals into a system of auxiliary-mode equations \cite{croy_propagation_2009, hu_communication_2010}.\\
Using the exactly solvable damped quantum harmonic oscillator (DQHO) as a benchmark, we propose an alternative approach to render featureless environments computationally tractable within the Schwinger-Keldysh formalism, without auxiliary equations. To this end, let us begin by analyzing the concrete difficulties that arise for the DQHO in the standard treatment.

\section{Schwinger-Keldysh field theory of the DQHO}\label{subsec:ohmically damped QHO}
Consider a single harmonic oscillator of mass $m=1$ and resonance frequency $\omega_0$ with the action
\begin{equation}
    S_\text{sys}  = \int_\mathcal{C}\frac{1}{2}\left(\left(\partial_t\varphi\right)^2 - \omega_0^2 \varphi^2\right) = \int_\mathcal{C}\frac{1}{2}\left(\varphi\, G_0\inv \varphi\right)\,.
\end{equation}
The integral over $\mathcal{C} = [t_0,\infty)\oplus (\infty,t_0]\oplus[t_0,t_0-i\beta)$ denotes the closed time path with the imaginary appendage to enforce a thermal equilibrium initial state, $\hat\rho_0 = \frac{e^{-\beta \hat H}}{\Tr e^{-\beta \hat H}}$ \cite{stefanucci_nonequilibrium_2013}. The generating functional is then given by
\begin{equation}
    Z[J] = \int D\varphi \, e^{i\left(S[\varphi]-\varphi\cdot J \right)}\,.
\end{equation}
We couple the oscillator to an environment that we assume is in thermal equilibrium. The microscopic details of this environment are unimportant for the discussion that follows. One may consider, e.g., the Caldeira-Leggett model \cite{caldeira_influence_1981, caldeira_path_1983}, an infinite reservoir of harmonic oscillators of varying frequencies with a bilinear coupling to the system (see \ref{sec:Caldeira-Legget}). If the reservoir and the system-reservoir interaction are simple enough, the environment can be integrated out exactly. In general, if the system couples linearly with arbitrarily strong coupling to an environment, which is itself not strongly interacting, the correction to the system dynamics due to the system-reservoir interaction can be captured perturbatively by an embedding self-energy that dresses the propagator,
\begin{equation}
    G\inv (t_1,t_2) = G\inv_0 (t_1,t_2) - \Sigma_ \emb (t_1,t_2) \,.
\end{equation}
The equation of motion of the field expectation value $\phi = \expval{\varphi^\cl}$ is modified accordingly as
\begin{equation}\label{eq:field_expval_eom}
    J^\cl(t) = \partial_t^2 \phi(t) + \omega_0^2 \phi(t) + \int_{t_0}^\infty dt' \Sigma_ \emb^\ret(t,t') \phi(t')\,.
\end{equation}
where we made use of the classical-quantum basis \cite{kamenev_field_2011} for the sources and fields i.e. $\varphi^\cl = \frac{1}{2}\left(\varphi^+ + \varphi^-\right)$ and $\varphi^\q = \varphi^+ - \varphi^-$.
Since the bath is in thermal equilibrium, the embedding self-energy is subject to the fluctuation-dissipation theorem (FDT). It is therefore completely characterized by its spectral function (sometimes called rate-function for self-energies \cite{stefanucci_nonequilibrium_2013}), $\Gamma_ \emb(\omega) = -2 \Im \Sigma_ \emb^\ret(\omega)$. The most prominent such spectrum is the ohmic one,
    \begin{equation}
    \Gamma_{\text{Ohm}}(\omega) = 2\gamma \omega \,.
\end{equation}
A range of different reservoirs and system-reservoir couplings can give rise to this effective model \cite{ford_quantum_1988,weiss_quantum_1999}, hence our indifference to the reservoir's microscopics. Importantly, the continuity of the spectrum eliminates any finite recurrence time and therefore enables truly dissipative dynamics \cite{barnett_revisiting_2023}.
The main reason for this particular spectrum's popularity is that the field expectation value reproduces exactly the classical damped harmonic oscillator EOM, 
\begin{equation}\label{classical_EOM}
    \Ddot{\phi} + \gamma \dot \phi + \bar\omega_0^2\phi = 0\,.
\end{equation}
Taking into account thermal but not quantum fluctuations (see \ref{subsec:classical_limit}) one recovers the classical Langevin equation \cite{kamenev_field_2011},
\begin{equation}\label{eq:Langevin}
    \Ddot{\phi} + \gamma \dot \phi + \bar\omega_0^2\phi = \xi(t)\text{ with }\expval{\xi(t)\xi(t')} = 2\gamma T \delta(t-t')
\end{equation}
where $\xi(t)$ is a Hubbard-Stratonovich field obeying a white noise distribution and $\bar\omega_0$ is the renormalized resonance frequency in the sense of the next paragraph.
In fact, the retarded propagators of the ohmically damped QHO and the classical DHO are identical without any approximations \cite{kamenev_field_2011}.
More general bath spectra are typically classified by their low frequency scaling $\Gamma(\omega\ll \omega_c) \sim \omega^s$ as sub-ohmic $s<1$, ohmic $s=1$, and super/supra-ohmic $s>1$, $s$ being the so-called ohmicity \cite{weiss_quantum_1999}.\\

It was recognized early on that the ohmic bath leads to at least two divergences caused by the unboundedness of the rate function: a linearly divergent correction to the resonance frequency \cite{caldeira_path_1983} and a logarithmically diverging expectation value of the momentum variance $\expval{\pi^2}$ \cite{grabert_quantum_1984}. To deal with these divergences, it is necessary to regularize the spectrum by introducing a cutoff scale \cite{weiss_quantum_1999}, e.g., via
\begin{equation}\label{eq:regulated_rate_function}
    \Gamma(\omega) = 2\gamma \omega \, e^{-\frac{\abs{\omega}}{\omega_c}}\,,
\end{equation}
such that the ohmic spectrum is obtained in the limit $\omega_c \to \infty$. 
In real systems, $\omega_c$ is always finite and set by a combination of the width of the bath density of states and a decay behavior of the system-reservoir coupling. Frequently, $\omega_c \gg \omega_0$ is assumed, and approximations based on this fact are used. In particular, the low frequency linear scaling is independent of the function that is used for the high-frequency regularization, $e^{-\frac{\abs{\omega}}{\omega_c}}$ in our case, a sharp cutoff $\theta(\omega_c-\abs{\omega})$ or e.g. the Drude regularization $\frac{\omega_c^2}{\omega^2 + \omega_c^2}$.\\
Given a reservoir with the rate-function \eqref{eq:regulated_rate_function}, the QHO's resonance frequency is renormalized \cite{caldeira_path_1983} according to
\begin{equation}\label{eq:mass_renormalization}
    \omega_0^2\rightarrow \omega_0^2 - \frac{2\gamma \omega_c}{\pi}\,.
\end{equation}
which is sometimes called “Lamb shift” due to its similar origin\footnote{The atomic levels get renormalized because they couple to a ($T\approx 0$) bath of photons. However, the actual Lamb shift between the $2s_{1/2}$ and $2p_{1/2}$ levels of hydrogen occurs due to the \textit{difference} of their coupling strengths to the bath, such that they do not get renormalized by the same amount and therefore become non-degenerate.}. For completeness, we note that the correction in \eqref{eq:mass_renormalization} is just the leading (and the only diverging) term of $\Re \Sigma^\ret_ \emb(\omega)$. We derive the subleading terms which play a role for $\omega_c\gtrsim \omega_0$ in Appendix \ref{sec:mass renorm}.
In principle, given a finite bath responsiveness $\omega_c$, both the bare and renormalized frequency retain a physical significance \cite{barnett_revisiting_2023}. On time scales $\delta t$ much shorter than $1/\omega_c$, expanding \eqref{eq:field_expval_eom} with the self-energy given by \eqref{eq:self-energy_A} the oscillator evolves according to 
\begin{equation}
    \Ddot{\phi}(t) + \omega_0^2 \phi(t) + \mathcal{O}(\delta t^2)= 0
\end{equation}
without damping, at the bare resonance frequency $\omega_0$. Intuitively, the environment cannot respond quickly enough to influence the system. For times much longer than $1/\omega_c$ the EOM is given by \eqref{classical_EOM}, that is, velocity-damped and with a renormalized frequency $\bar \omega_0 = \sqrt{\omega_0^2 - \frac{2\gamma\omega_c}{\pi}}$ resulting in a damped oscillation of frequency $\omega_\gamma = \sqrt{\bar\omega_0^2-\frac{\gamma^2}{4}}$. In practice, relatively small $\omega_c$ and large $\gamma$ are required for the difference in dynamics not to be obscured by the uncertainty relation between time and frequency. If one is not interested in or does not have access to the fast timescale, distinguishing the renormalized resonance frequency with a bar is pointless. Instead, one may consider the limit $\omega_c\to\infty$ relative to the timescales of interest. Then, there is also formally no way to isolate the bare resonance frequency in the measurement of any observable. One can include a counter-term in the action by defining the bare parameter $\omega_0 \to \omega_0 + \frac{2\gamma \omega_c}{\pi}$ to cancel the correction and consider $\omega_0$ as the renormalized frequency. This is typically done in the literature \cite{caldeira_path_1983}, and we will proceed in the same way in the following.\\

The second divergence occurs in the momentum variance. In thermal equilibrium, it is given by
\begin{equation}
    \expval{\pi^2} = T\sum_{n=-\infty}^\infty \frac{\omega_0^2 + \gamma \abs{\omega_n}e^{-\frac{\abs{\omega_n}}{\omega_c}}}{\omega_n^2 + \omega_0^2 + \gamma \abs{\omega_n} e^{-\frac{\abs{\omega_n}}{\omega_c}}}\,,
\end{equation}
with $\omega_n$ the bosonic Matsubara frequencies, which diverges logarithmically for $\omega_c \to \infty$ \cite{grabert_quantum_1984}.
There has been at least one attempt to renormalize the second divergence as well \cite{agon_divergences_2018}. Working within the operator formalism, their approach was to renormalize the $\hat X \hat P$ and $\hat P^2$ operators, yielding closed, finite EOMs for these operators. We find it difficult to assign a physical interpretation to their renormalized operators. The point of view we adopt is that the $\omega_c$-dependence is physical but renormalized by a finite temporal resolution present in experiments and in numerical simulations on a grid (see \ref{sec:Renormalization}). Although the exponential regulator is only a rough proxy for more realistic decay behavior and $\omega_c$ therefore has a large uncertainty, the logarithmic divergence arises for arbitrary regularizations due to dimensional reasons. It has the redeeming quality of low $\omega_c$-uncertainty propagation.\\

\subsection{(Non-)markovianity of the DQHO}
One of the strengths of the Schwinger-Keldysh formalism is its ability to treat systems that do not admit to the Born-Markov approximation. This approximation entails essentially memoryless time-evolution, which is used to derive the popular Redfield and Lindblad equations (the latter requiring an additional secular approximation). For comparison, we briefly review these concepts and classify the possible parameter/scale configurations of the DQHO.\\
The most general notion of Markovian evolution of an open quantum system is given by the equation \cite{gonzalez-ballestero_tutorial_2024}
\begin{equation}
    \dot\rho(t) = \mathcal{L}_t[\rho(t)]
\end{equation}
where $\mathcal{L}_t$ is some time-local super-operator (not necessarily a Lindbladian), hence the change of the density matrix depends only on its current state. In contrast, a general non-Markovian time evolution of the density matrix is of the form
\begin{equation}
    \dot\rho(t) = \int_0^t dt'\, \mathcal{K}(t,t')[\rho(t')]\,.
\end{equation}
The emergence of time-nonlocality is a natural consequence of integrating out the environment since this imposes convolutions with bath correlators, i.e., two-time functions. The Markovian equation is trivially recovered for $\mathcal{K}(t,t')\propto \delta(t-t')$. This is a valid approximation if the bath correlation decay rate, $\tau_\text{mem}\inv$, is larger than all system scales. In this case $\rho(t-\tau)\approx \rho(t)$ over the support of $\mathcal{K}(t,t-\tau)$. The Born-Markov approximation may be viewed as a more subtle extension of the above. The assumption that $\tau_\text{mem}$ is the shortest time-scale is relaxed only to require that it be faster than the dissipation rate of the system, $\tau_\text{mem}\ll \tau_\text{diss}$, where $\tau_\text{diss} \sim \gamma\inv$. Crucially, the scale of the coherent dynamics (phase oscillations with $e^{i\omega_0 t}$) is allowed to be faster than the memory scale. In the interaction picture w.r.t. the unitary part of $\mathcal{K}$, $\rho_I(t)$ is constant under coherent evolution and only varies on the dissipation time-scale $\tau_\text{diss}$. In exchange, the coherent dynamics and possible interferences thereof over the environment's memory depth are absorbed into the interaction picture super-operator $\mathcal{K}_I(t,t')$. Since, by assumption, $\rho_I$ does not significantly vary over $\tau_\text{mem}$, the Markov approximation can be performed in the interaction picture. A more detailed and pedagogical derivation of these approximations can be found in \cite{gonzalez-ballestero_tutorial_2024}.\\
Let us now consider how these concepts apply to the DQHO. In total, there are four relevant energy/time scales: The damped oscillator's (renormalized) frequency $\omega_0$ (i.e., the coherent evolution rate), its damping rate $\gamma$, the bath temperature $T$, and finally the bath spectrum cutoff scale $\omega_c$. A priori, it is not obvious how the two scales $T$ and $\omega_c$ relate to the abstract scale $\tau_\text{mem}$. However, it will be shown in \ref{sec:nonequilibrium} that $\tau_\text{mem} = \text{max}\left\{\omega_c\inv, T\inv\right\}$. In the following, we focus on the parameter regime in which $\omega_c$ is the largest among all scales (the wide-band limit). If also the temperature is larger than the remaining two scales, i.e., $T,\omega_c \gg \omega_\gamma,\gamma$, we are in the strictly Markovian regime and recover the classical damped oscillator (see \ref{subsec:classical_limit}). Hence, the quantum mechanically interesting constellations of the remaining scales have temperatures on the same order or smaller than $\omega_\gamma$ and $\gamma$. In particular, we can distinguish the cold regime, $\omega_\gamma \gg T \gg \gamma$, with relevant quantum fluctuations but (born-)markovian evolution (so long as we do not add external driving), and the ultra-cold regime, $\omega_\gamma,\gamma \gg T$. In this last regime, particularly for strong damping, the system exhibits markedly different behaviour from the classical and weakly damped regimes. The displacement and momentum uncertainties show strong squeezing (see \ref{subsec:uncertainty_relations}) and correlations follow a power-law instead of exponential scaling \cite{haake_strong_1985}. Lastly, $2\omega_0 \gtrless \gamma$ determines whether the oscillator is under- or overdamped, as in classical theory, and, for simplicity, we focus on the underdamped case. While Keldysh/NEGF makes no approximation, and is therefore formally valid in all parameter regimes, naive time-stepping implementations are numerically prohibitive for a large separation of scales between $\omega_c$ and $\omega_0$ since the resulting differential equations become stiff. In \ref{sec:Conclusion}, we propose a way to remove this limitation by explicitly using the separation of scales to our advantage. As a consequence, it becomes feasible to use the Schwinger-Keldysh formalism for nonequilibrium implementations of high-bandwidth but very cold/coherent environments.

%% file: sections/Equilibrium.tex
\section{Equilibrium}\label{subsec:equilibrium results}
Due to the problem's dissipative nature, we expect that for $ t\to\infty$ the system thermalizes at the bath temperature $T_B$, irrespective of the initial state. Therefore, it is useful to analyze the equilibrium solutions obtained within the imaginary-time formalism to determine the asymptotic values of any real-time evolution of the system. For the strictly ohmic limit, $s=1$ and $\omega_c \to \infty$, the equilibrium solution for $G$ is even known analytically \cite{weiss_quantum_1999}. Let us clarify our conventions and notation for the Keldysh components. The propagator is defined as G(1,2) = $-i \expval{\mathcal{T}_\mathcal{C}\varphi(1)\varphi(2)}$ and the separation into a real antisymmetric and imaginary symmetric part reads
\begin{gather}
    G^A(1,2) = -i\expval{[\varphi(1),\varphi(2)]}\\
    G^S(1,2) = -\frac{i}{2}\expval{\{\varphi(1),\varphi(2)\}}  + i \phi(1)\phi(2)
\end{gather}
where the last term is only necessary for non-vanishing field expectation values to obtain the connected correlation function (cumulant). Using this decomposition, the propagator is given by
\begin{equation}
    G(1,2) = G^S(1,2) + \frac{1}{2}\mathrm{sgn}_\mathcal{C}(t_1-t_2) G^A(1,2)
\end{equation}
with $\text{sgn}_\mathcal{C}$ the sign operator with respect to contour time ordering. All real-time Keldysh components can be expressed in terms of these two, i.e.
\begin{align}
    G^{\ret/\adv}(1,2) &= \pm \theta(\pm(t_1-t_2)) G^A(1,2)\,,\\
    G^\gtrless(1,2) &= G^S(1,2) \pm \frac{1}{2} G^A(1,2)\,.
\end{align}
In equilibrium, the antisymmetric and symmetric components can be obtained via the FDT and Fourier transformation from the spectral function \cite{stefanucci_nonequilibrium_2013}, 
\begin{equation}
    A(\omega) = i\left[G^>(\omega)-G^<(\omega)\right]\,,
\end{equation}
given by
\begin{equation}
    A(\omega) = \frac{2\gamma \omega }{(\omega^2-\omega_0^2)^2 + (\gamma \omega)^2}\,,
\end{equation}
to be
\begin{equation}
    G^A(t) = - \frac{1}{\omega_\gamma}\sin(\omega_\gamma t)e^{-\frac{\gamma}{2}\abs{t}}
\end{equation}
and
\begin{equation}
    G^S(t) = \frac{1}{2}\int_\omega e^{-i\omega t} \coth\left(\frac{\omega}{2 T}\right) G^A(\omega)\,.
\end{equation}
The integral is solved via contour integration. Following Weiss \cite{weiss_quantum_1999} we partition $G^S(t)$ into a contribution due to the poles of the spectral function at $\omega_\gamma \pm i\frac{\gamma}{2}$ and $-\omega_\gamma \pm i\frac{\gamma}{2}$ and a contribution due to the poles at the Matsubara frequencies $i\omega_n = 2\pi i T n,$ for $n\in \mathbb{Z}$.
\begin{widetext}
    \begin{equation}\label{eq:symmetric_solution}
    G^S(t) = G^S_1(t) + G^S_2(t) = -i \frac{1}{2\omega_\gamma}\frac{\sinh(\omega_\gamma \beta)\cos(\omega_\gamma t)+\sin(\frac{\gamma}{2}\beta)\sin(\omega_\gamma\abs{t})}{\cosh(\omega_\gamma \beta)-\cos(\frac{\gamma}{2}\beta)}e^{-\frac{\gamma}{2}\abs{t}} + i 2\gamma T \sum_{n=1}^\infty \frac{\omega_n e^{-\omega_n \abs{t}}}{(\omega_n^2+\omega_0^2)^2-\gamma^2\omega_n^2}
\end{equation}
\end{widetext}
The equilibrium correlators share a few general properties with the solutions for any other state. The first derivative of the antisymmetric correlator at equal times is fixed by the canonical commutation relations (CCR)\footnote{In the $\omega_c \to \infty$ limit, the embedding self-energy in the effective action contains a time derivative. This leads, in principle, to a modified canonical momentum,
\begin{equation}
    \pi(t) = \frac{\partial\mathcal{L}(t)}{\partial\partial_t\varphi(t)} = \partial_t\varphi(t) - \gamma \varphi(t)\,,
\end{equation}
distinct from the kinematic momentum. However, the CCR $[\hat\varphi(t),\hat\pi(t)] = i$ still implies $[\partial_t\hat\varphi(t),\hat\varphi(t)] = -i$, since the modification clearly commutes with $\hat \varphi$. Hence, the equal time relations survive the $\omega_c\to\infty$ limit.} to 
\begin{equation}\label{eq:G_A CCR}
    \partial_{t_1} G^A(t_1,t_2)\Big\vert_{t_1=t_2} = -1\,.
\end{equation} 
The second derivative of $G^A$ has a discontinuity on the time-diagonal given by
\begin{equation}\label{eq:discontinuity_of_G_A}
    \partial_{t_1}^2 G^A(t_1,t_2) = \begin{cases}
        -\gamma,\quad &\text{for } t_1-t_2=0^-\\
        0, \quad &\text{for } t_1-t_2=0\\
        \gamma,\quad &\text{for } t_1-t_2=0^+
    \end{cases}\,,
\end{equation}
where the exact equality to zero at equal times follows from antisymmetry. It will become clear in the following that this discontinuity appears regardless of the state in the ohmic limit because it follows directly from the Kadanoff-Baym equation,
\begin{equation}
    \partial_{t_1}^2 G^A_{1,2} = -\omega_0^2 G^A_{1,2} - \gamma\, \sgn(t_1-t_2) \partial_{t_1} G^A_{1,2}\,
\end{equation} 
due to the instantaneous damping term and \eqref{eq:G_A CCR}.\\
The second derivative of $G^S$ has a singularity at equal times. This is exactly the logarithmic divergence of the momentum variance, since in the steady state
\begin{equation}
    \partial_t^2 G^S(t)\big\vert_{t=0} = - \partial_{t_1}\partial_{t_2} G^S(t_1-t_2)\big\vert_{t_1=t_2}=i\expval{\pi^2}
\end{equation}
It will be shown later that this is a state-independent property. Furthermore, since the singularity is logarithmic, it is integrable and therefore $\partial_{t_1} G^S(t_1,t_2)$ is continuous with an infinite slope at equal time (like e.g., $\sgn(x)\sqrt{\abs{x}}$ at $x=0$).

\subsection{Uncertainty relations}\label{subsec:uncertainty_relations}
\begin{figure*}
    \centering
    \includegraphics[width=0.99\linewidth]{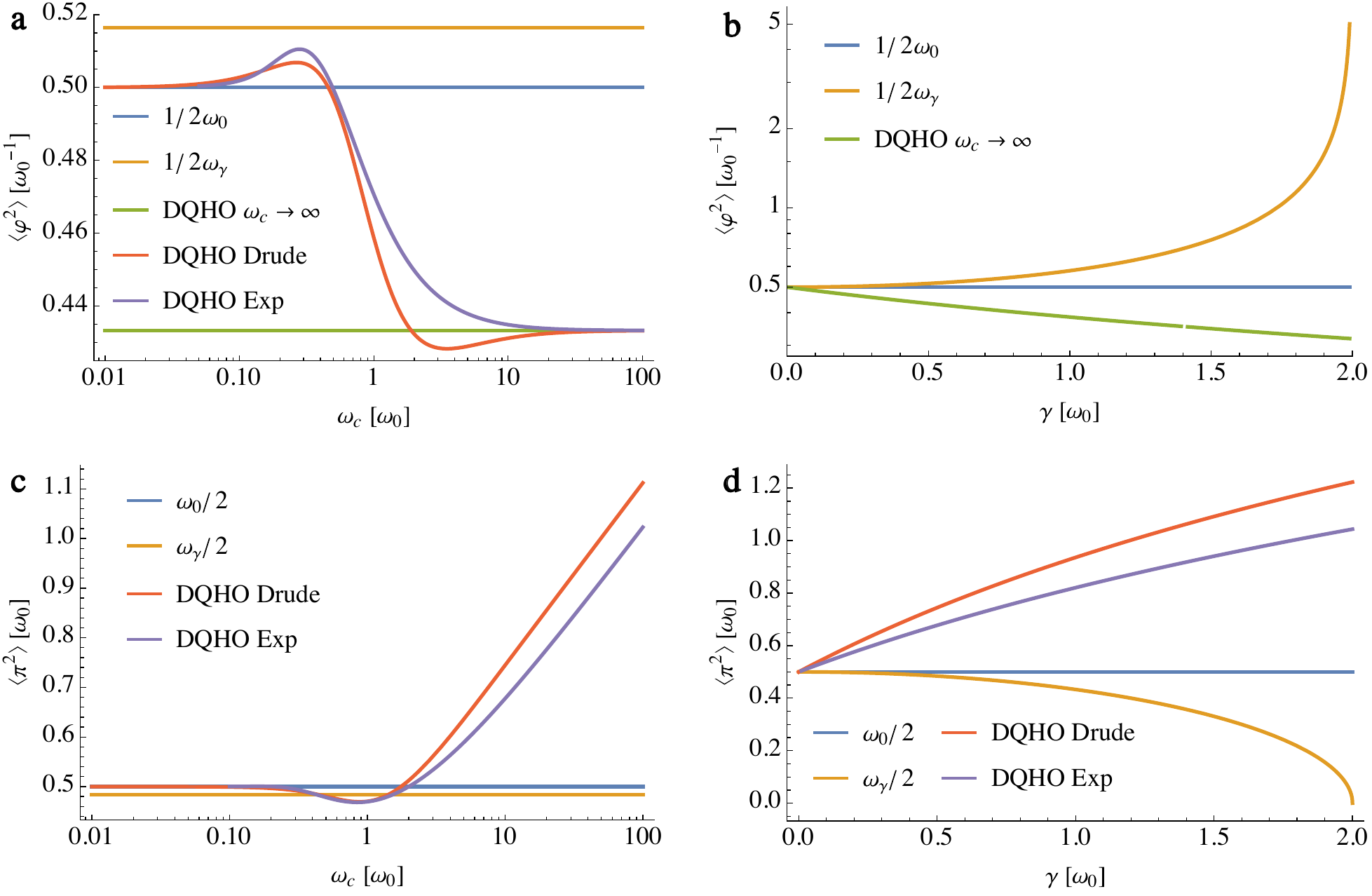}
    \caption{(a): Comparison of the undamped and damped QHO amplitude variances for $\gamma/\omega_0 = 0.5$. The non-constant graphs, (red) for Drude regularization and (violet) for exponential regularization, show that the amplitude variance as a function of $\omega_c$ interpolates between the undamped oscillator (blue) and the ohmic limit \eqref{eq:field_variance_ohmic} (green). For comparison, we also present the value for an undamped oscillator with the same resonance frequency, $\omega_\gamma$, as the damped one (yellow). (b): Scaling of the amplitude variance with damping strength $\gamma$ (green) compared to the undamped oscillator with the same resonance frequency (yellow) diverging at $\gamma = 2\omega_0$. (c)Comparison of the undamped and damped QHO momentum variances for $\gamma/\omega_0 = 0.5$ as a function of $\omega_c$. We compare the Drude regularization (red) with the exponential regularization (violet). Both converge to the isolated oscillator for $\omega_c \to 0$ and scale logarithmically for $\omega_c \to \infty$. (d): We show the momentum variance of the damped QHO at $\omega_c = 10\omega_0$ compared to the isolated QHO, both with the original frequency $\omega_0$ and with the same resonance frequency as the damped QHO, $\omega_\gamma$. We also compare the Drude regularization (red) with the exponential regularization (violet), both of which exhibit the same qualitative behavior: monotonically increasing as a function of $\gamma$.} 
    \label{fig:variances}
\end{figure*}
The equilibrium state of the QHO has well-known amplitude and momentum variances $\Var_T(X) = \frac{1}{2\omega_0}\coth(\frac{\omega_0}{2T})$ and $\Var_T(P) = \frac{\omega_0}{2}\coth(\frac{\omega_0}{2T})$. In the high-temperature limit, one can approximate $\coth(\frac{\omega_0}{2T})\approx \frac{2T}{\omega_0}$ and recover the classical equipartition theorem (in natural units and with $m=1$),
\begin{equation}\label{eq: equipartition uncertainty}
    \expval{X^2}_T = \frac{T}{\omega_0^2}\,,\qquad \expval{P^2}_T = T\,.
\end{equation}
On the other hand, for $\coth(\frac{\omega_0}{2T})\overset{T\to 0}{\longrightarrow} 1$ the position-momentum uncertainty relation is saturated, i.e.,
\begin{equation}
    \Var_0(X) \Var_0(P) = \frac{1}{4}\,.
\end{equation}
Interestingly, the uncertainty relations of the damped QHO are not given by replacing $\omega_0 \to \omega_\gamma$ but depend non-trivially both on the damping strength $\gamma$ and the bath responsiveness $\omega_c$ \cite{grabert_quantum_1984,grabert_quantum_1988}. In the ohmic limit, the amplitude variance is obtained using \eqref{eq:symmetric_solution} at $t=0$. Its temperature dependence is of minor interest to us, and we refer to \cite{grabert_quantum_1988,weiss_quantum_1999}, where it is discussed in detail. In the low-temperature limit and underdamped regime ($\gamma < 2\omega_0$), the result may be compactly written as
\begin{equation}\label{eq:field_variance_ohmic}
\begin{split}
    \expval{\varphi^2}_0 &= \frac{\pi + 2\text{ ArcCot}\left(\frac{\gamma \omega_\gamma}{\omega_\gamma^2 - (\gamma/2)^2}\right)}{4\pi\omega_\gamma}\\
    &\approx \frac{1}{2\omega_\gamma}\left( 1 - \frac{\gamma}{\pi \omega_\gamma} + \mathcal{O}\left(\frac{\gamma^2}{\omega_\gamma^2}\right)\right)\,.
\end{split}
\end{equation}
It is worth mentioning that not only is this smaller than the naive replacement $1/2\omega_\gamma$, but also smaller than $1/2\omega_0$ and decreasing for stronger damping (see fig. \ref{fig:variances}).
For finite $\omega_c$, neither $G^S(t)$ nor $\expval{\varphi^2}$ can be expressed analytically using the exponential regulator. However, we show its dependence on $\omega_c$ graphically in Figure \ref{fig:variances}.
As was already mentioned, unlike the field variance $\expval{\varphi^2}$, the momentum variance $\expval{\pi^2}$ does not converge for $\omega_c\to\infty$. One may write \cite{grabert_quantum_1984}
\begin{equation}
    \expval{\pi^2} = \omega_0^2 \expval{\varphi^2} + \Delta
\end{equation}
For the momentum variance, the temperature dependence is more interesting. In the regime $\omega_c \gg T$ but $\omega_\gamma \ll T$ one finds an asymptotic scaling $\Delta \sim \gamma\log(\omega_c/T)$, whereas if also $\omega_\gamma\gg T$, $\Delta \sim \gamma\log(\omega_c/\omega_\gamma)$ \cite{grabert_quantum_1984}. On the other hand, for $T\gg \omega_c$, the correction approaches zero as $\Delta \sim \omega_c/T$ and the classical result \eqref{eq: equipartition uncertainty} is recovered once more \cite{grabert_quantum_1984}.

%% file: sections/Nonequilibrium.tex
\section{\label{sec:nonequilibrium}Nonequilibrium}
\subsection{Kadanoff-Baym equations}\label{subsec:KBEs}
\begin{figure}
    \centering
    \includegraphics[width=\linewidth]{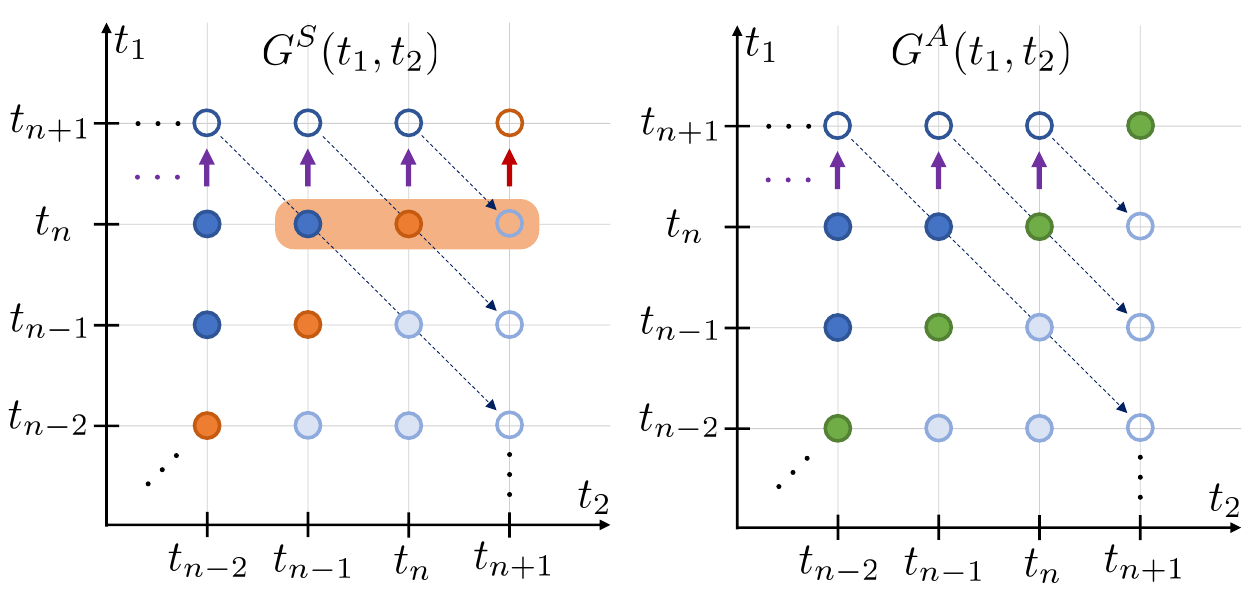}
    \caption{Illustration of the time-stepping algorithm in the time-plane for symmetric (left) and antisymmetric (right) correlators (in the style of \cite{schuler_time-dependent_2016}). Each step from $t_n\to t_{n+1}$ consists of two sub-steps. First, evolve $G(t_n,t_k) \to G(t_{n+1},t_k)$ for all $k\leq n$ (violet arrows). The lower triangle contains redundant information, due to the respective symmetry,  $G^{S/A}(t_k,t_{n+1}) = \pm G^{S/A}(t_{n+1},t_k)$ (dashed arrows). For the antisymmetric correlator, the diagonal vanishes exactly, so no further action is required (green dots). For the symmetric correlator, the diagonal elements capture the amplitude variance; on it, the KBE diverges logarithmically with $\omega_c$ (orange dots). It is obtained via the additional step $G^S(t_n,t_{n+1}) \to G^S(t_{n+1},t_{n+1})$ (red arrow).}.
    \label{fig:time_stepping}
\end{figure}

The Kadanoff-Baym equations (KBEs) are the out-of-equilibrium analog of the Dyson equation. The left/right KBE is the equation of motion of the propagator, $G(t_1,t_2)$, with respect to $t_1$/$t_2$, that is,
\begin{equation}\label{eq:Kadanoff-Baym_general}
\begin{split}
    G_0\inv \cdot G &= 1 + \Sigma \cdot G\\
    G \cdot G_0\inv &= 1 + G \cdot \Sigma 
\end{split}
\end{equation}
where we use the shorthand $A\cdot B = \int_\mathcal{C}dt' A(t_1,t') B(t',t_2)$ on the CTP, and the self-energy is in general a functional of both the propagator and the field. Multiplying from the left/right by $ G_0$ yields the usual Dyson equation. However, \eqref{eq:Kadanoff-Baym_general} formulates the same relation in terms of an integro-differential initial value problem, which is more suitable for the nonequilibrium scenario. This becomes clearer when writing out the inverse bare propagator,
\begin{align}\label{eq: KBEs on C}
-(\partial_{t_1}^2 + \omega_0^2)G(t_1,t_2) &= \delta_\mathcal{C}(t_1,t_2)  + (\Sigma\cdot G)(t_1,t_2)\\
-(\partial_{t_2}^2 + \omega_0^2)G(t_1,t_2) &= \delta_\mathcal{C}(t_1,t_2)  + (G\cdot\Sigma) (t_1,t_2)
\end{align}
In contrast to other partial (integro-)differential equations, the KBEs can be significantly simplified by noting that both variables, being time, share the same domain and causality.
 To make the equation easier to handle, the equations \eqref{eq:Kadanoff-Baym_general}, formulated in terms of functions on the contour $\mathcal{C}$, are replaced by functions defined on real intervals using the Langreth rules \cite{stefanucci_nonequilibrium_2013, berges_introduction_2004}. The left-sided (and analogously the right-sided) KBEs of the (anti-)symmetric- and right correlators are given by
\begin{widetext}
\begin{align}\label{eq:antisymkbe}
    G_0\inv G^A (t_1,t_2) &= \int_{t_2}^{t_1}\Sigma^A (t_1,t') G^A(t',t_2)\\
    \label{eq:symkbe}
    G_0\inv G^S (t_1,t_2) &= \int_{t_0}^{t_1}\Sigma^A (t_1,t') G^S(t',t_2) - \int_{t_0}^{t_2} \Sigma^S(t_1,t') G^A(t',t_2) - i\int_0^\beta d\tau \Sigma^\urcorner(t_1,\tau) G^\ulcorner (\tau,t_2)\\
    \label{eq:rightkbe}
    G_0\inv G^\urcorner (t,\tau) &= \int_{t_0}^{t} dt' \Sigma^A(t,t')G^\urcorner(t',\tau) - i \int_0^\beta d\tau'\Sigma^\urcorner(t,\tau')G^M(\tau',\tau)
\end{align}
\end{widetext}
For the bosonic field correlators, $G^\urcorner(t,\tau) = G^\ulcorner(\tau,t)$, such that the EOM for the right component also provides the left component that enters \cref{eq:symkbe}, closing \cref{eq:antisymkbe,eq:symkbe,eq:rightkbe}.
One can exploit the symmetries of $G^{A/S}$ to rely on only the left (or right) KBEs and therefore evolve in only one of the time directions (see Fig \ref{fig:time_stepping}). Effectively, the system \eqref{eq: KBEs on C} of partial integro-differential equations is replaced by three coupled ordinary integro-differential equation families, which can be solved via a time-stepping algorithm. Treating $G^A$ and $G^S$ separately emphasizes an extra degree of freedom, containing information about the occupation of one-particle excitations, compared to the equilibrium theory. In the latter, the occupation is fixed by the equilibrium statistics, given the single particle spectrum contained in $G^A$. Accordingly, the statistical/occupational degree of freedom in $G^S$ is lost via the fluctuation dissipation relation (FDR).\\
The embedding self-energy does not depend on the system. In the non-interacting DQHO model, the KBE for the antisymmetric correlator can be solved independently. It encodes the single-particle spectrum of the system and is therefore sensitive to changes in the Lagrangian. More generally, interactions generate correlation self-energies in addition to the embedding component, i.e.,
\begin{equation}
    \Sigma = \Sigma_ \emb + \Sigma_c\,,
\end{equation}
which couples \eqref{eq:antisymkbe} to the other KBEs via the correlation self-energy's functional dependencies.\\
We assume that the environment is much larger than the system and therefore remains in thermal equilibrium regardless of what happens within the system. This assumption allows us to use the FDR to relate the antisymmetric and symmetric components of the embedding self-energy. Accordingly, they are given by
\begin{align}
    \Sigma^A_ \emb(\omega) &= -i\Gamma(\omega)\\
    \Sigma^S_ \emb(\omega) &= \frac{1}{2}\coth(\frac{\omega}{2T_B})\Sigma^A_ \emb(\omega)\,. \label{eq:Sigma_S_freq}
\end{align}
It is a standard result to show that if the system evolves to a steady state, it is the thermal state \cite{stefanucci_nonequilibrium_2013}. To this end, we assume that for large times the initial correlations $G^\urcorner,\Sigma^\urcorner$ have decayed, and the correlators become translation invariant. Then we can act with $G_0^\ret$ from the left on \eqref{eq:antisymkbe} and rearrange the equation for $G^S$ to read
\begin{equation}
    \left(G_0\inv - \Sigma^\ret\right)\cdot G^S = \Sigma^S\cdot G^\adv\,,
\end{equation}
where we identify the bracket on the left as the inverse of $G^\ret$ such that it can be brought to the other side, giving
\begin{align}
    G^\ret &= G_0^\ret\cdot\Sigma^\ret\cdot G^\ret\\
    G^S &= G^\ret\cdot\Sigma^S\cdot G^\adv\,.
\end{align}
Now, translation invariance can be used to perform a Fourier transform, thereby trading convolutions in time for products in frequency. The first equation is solved by
\begin{equation}
    G^\ret(\omega) = \frac{1}{\omega^2-\omega_0^2 - \Sigma^\ret(\omega)}
\end{equation}
Physically, the crucial step is to use that the bath is in equilibrium by inserting the FDR for the embedding self-energy \eqref{eq:Sigma_S_freq} such that one can use $\Sigma^A(\omega) = \Sigma^\ret(\omega) - \Sigma^\adv(\omega)$ and \begin{equation}
    G^\ret\left(\Sigma^\ret-\Sigma^\adv\right)G^\adv = G^\ret - G^\adv
\end{equation}
to find
\begin{equation}
    G^S(\omega) = \frac{1}{2} \coth\left(\frac{\omega}{2T_B}\right)\left(G^\ret(\omega) - G^\adv(\omega)\right)
\end{equation}
which shows that the system correlator inherits the FDR from the bath self-energy and therefore the system satisfies the KMS condition, indicating that it thermalizes at the bath's temperature.

\subsection{Interpreting the embedding self-energy}\label{subsec:embedding self energy}
For the generalized low-energy bath spectrum
\begin{equation}\label{eq:generalized_rate_function}
    \Gamma_s(\omega) = 2\gamma \omega \left(\frac{\abs{\omega}}{\omega_c}\right)^{s-1} e^{-\frac{\abs{\omega}}{\omega_c}}
\end{equation}
of ohmicity $s$, the Fourier transformations of the embedding self-energy components \eqref{eq:Sigma_S_freq} (needed for a Keldysh implementation) can be obtained analytically. For the ohmic ($s=1$) case we find (see appendix \ref{sec:Fourier_derivation} for the Fourier transform and appendix \ref{sec:HFP and CPV identities} for the limit)
\begin{widetext}
    \begin{align}\label{eq:self-energy_A}
    \Sigma^A_ \emb(t_1,t_2) &= - 2\gamma \frac{2(t_1-t_2)\omega_c\inv}{\pi\left(\omega_c^{-2}+(t_1-t_2)^2\right)^2} \quad \overset{\omega_c\to\infty}{\longrightarrow}\quad 2\gamma \delta'(t_1-t_2)\\
    \Sigma^S_ \emb(t_1,t_2) &= i\gamma \pi T^2 \frac{2\cos\left(\frac{2\pi T}{\omega_c}\right)\cosh\left(2\pi T(t_1-t_2)\right) - 2}{\left(\cos\left(\frac{2\pi T}{\omega_c}\right)-\cosh(2\pi T(t_1-t_2) )\right)^2} + R_{\omega_c}(t_1-t_2) \quad \overset{\omega_c\to\infty}{\longrightarrow}\quad \fp \left( \frac{i\gamma \pi T^2}{\sinh^2(\pi T (t_1-t_2))}\right)\label{eq:self-energy_S}
\end{align}
\end{widetext}
where the term $R_{\omega_c}(t)$ scales like $T/\omega_c$ for $\omega_c\to\infty$. We consider $\omega_c \gg T_B$ and neglect this term in the following. Similar-looking limits can be found in the literature \cite{grabert_quantum_1988,agarwal_initial_2024}, however, only without the "$\fp$" denoting the Hadamard finite part (see Appendix \ref{sec:HFP} for an introduction) of the expression.\\
Neglecting this distributional aspect is not permissible, since, e.g., the zero frequency limit
\begin{equation}
    \lim_{\omega\to 0} \Sigma^S_ \emb(\omega) = - 2i\gamma T
\end{equation}
would not agree with the Fourier transform of the time parametrized $\Sigma^S_ \emb(t)$ at zero frequency,
\begin{equation}
    \int_{-\infty}^\infty dt' \frac{i\gamma \pi T^2}{\sinh^2(\pi T t')} 
\end{equation}
which suffers from an IR-divergence \cite{kamenev_field_2011}.
In the general scenario \eqref{eq:generalized_rate_function}, let us use that on the real axis 
\begin{equation}
    \coth(\omega/2T) = \sgn(\omega) + 2\sgn(\omega) n_B(\abs{\omega})
\end{equation}
where $n_B$ denotes the Bose-Einstein distribution. We separate the fluctuation kernel into a vacuum and a finite temperature component,
\begin{equation}
    \Sigma^S_s(\omega) = \gamma \abs{\omega}^s + 2\gamma \abs{\omega}^s n_B(\abs{\omega})\,,
\end{equation}
where we suppressed the exponential regulator for now. Clearly, the thermal part is perfectly finite for $\omega\to\pm\infty$ even without introducing a regulating $\omega_c$ due to the decaying Bose-Einstein distribution. Only the vacuum part diverges, and its Fourier transform needs to be interpreted as the finite part 
\begin{equation}
    \mathcal{F}\left(\abs{\omega}^s\right) = \fp \frac{1}{\abs{t}^{s+1}}
\end{equation}
Thus, the generic behaviour of $\Sigma^S_s(t)$ is a distribution interpolating between the finite part of $1/\abs{t}^{s+1}$ and an exponentially decaying tail for $ t\gg T\inv$, just like we find for the ohmic case \eqref{eq:self-energy_S}.
We conclude that the formulation in terms of (nascent) Hadamard finite parts is the right framework for arbitrary ohmicity. We note also that the exponential regulator, $e^{-\abs{\omega}/\omega_c}$, allows us to derive higher ohmicity embeddings from lower ones via the differentiation
\begin{equation}
    \Sigma_s^S(t;\omega_c) = \omega_c^2\partial_{\omega_c} \Sigma_{s-1}^S(t;\omega_c)\,.
\end{equation}
For simplicity, we focus on the ohmic case $s=1$ in the following. We will come back to the general case in the concluding remarks.\\

 % He cured this rather pragmatically by adding a term $-\delta(t)\left(2i\gamma T_B + C\right)$ to an otherwise unregulated $\Sigma_ \emb^S(t)$, $C$ being a diverging constant equal to the integral $\int_{-\infty}^{\infty}\frac{i\gamma \pi T^2}{\sinh^2(\pi T t)} dt$. This is very similar in spirit (and served as inspiration) to the finite part formulation we propose. In fact, for $t_0\ll t_1 \ll t_2$, that is, far away from the time grid's borders and diagonal, they coincide. However, this is not enough to ensure finite KBEs. Using this regularization, the RHS still diverges on the diagonal since $G^A(t,t)$ evaluates to zero. This matter will become clearer in Section \ref{sec:separation of scales}.
\subsection{The classical limit}\label{subsec:classical_limit}
In contrast to a damped classical harmonic oscillator (DCHO), the DQHO's fluctuations generated by $\Sigma_ \emb^S$ do not share their memory-scale with the dissipation due to $\Sigma_ \emb^A$. While $\Sigma_ \emb^A$ is non-local on the scale $\omega_c\inv$, $\Sigma^S_ \emb$ only starts to decay over the larger of the two scales, $\omega_c\inv$ and $\beta_B$ \cite{hu_quantum_1992,lindenberg_statistical_1984}. Because of this, the DCHO becomes Markovian in the limit $\omega_c\to\infty$,  while the DQHO does not. It is intuitive that, for very high temperatures (compared to $\omega_0$), thermal fluctuations dominate over quantum fluctuations, thereby erasing any phase coherence, so that the dynamics becomes Markovian even for the DQHO. This intuition can be made more precise by considering the high temperature limit $\hbar \omega_0/k_B T_B \to 0$ \cite{ford_quantum_1987,ford_quantum_1988}. Contrary to the claim made in \cite{hu_quantum_1992}, the order of the limits $\omega_c\to\infty$ and $T\to\infty$ does not matter. Let us state the trivial order first: Using $\coth(\frac{\omega}{2T})\to\frac{2T}{\omega}$ for $T\to\infty$, the inverse Fourier transform of \cref{eq:Sigma_S_freq} gives
\begin{equation}\label{eq:trivial_order_classic_limit}
\begin{split}
    \Sigma^S(t_1,t_2) &= -i\int \frac{d\omega}{2\pi} 2\gamma T e^{-\abs{\omega}/\omega_c}e^{-i\omega(t_1-t_2)}\\
    &= -i2\gamma T \frac{\omega_c^{-1}}{\pi(\omega_c^{-2}+(t_1-t_2)^2)}\\
    &\overset{\omega_c\to\infty}{\longrightarrow}-i2\gamma T \delta(t_1-t_2)\,.
\end{split}
\end{equation}
Thanks to the finite part interpretation, \cref{eq:self-energy_S}, we can make sense of first taking $\omega_c\to\infty$, and then $T\to \infty$ just as well as the other way around. The memory kernel $\fp \sinh^{-2}(x)$ (up to normalization) satisfies the conditions (Appendix \ref{sec:delta_squared_derivation}) to become a nascent delta function. Let us define a dimensionless time $x = t \omega_\gamma $ such that the $\omega_c \to \infty$ limit of the symmetric embedding self-energy reads
\begin{equation}
    \Sigma^S_ \emb(t_1,t_2) = \fp \left( \frac{i\gamma \pi T^2}{\sinh^2(\pi T (x_1-x_2)/\omega_\gamma)}\right)\,.
\end{equation}
With $\varepsilon = \omega_\gamma/T$ we find for $t_0<t_1<t_2$
\begin{widetext}
    \begin{equation}
    \ddashint_{t_0}^{t_2}\frac{i\gamma \pi T^2}{\sinh^2(\pi T (t'-t_1))} f(t') dt' = \ddashint_{x_0}^{x_2}-\frac{2i\gamma T}{\varepsilon}K\left(\frac{x'- x_1}{\varepsilon}\right) f(x'/\omega_\gamma) dx' \overset{\varepsilon\to 0}{\longrightarrow} - 2i\gamma T f(t_1)
\end{equation}
\end{widetext}
where we used the result
\begin{equation}
    \ddashint_{-\infty}^{\infty}\frac{i\gamma \pi T_B^2}{\sinh^2(\pi T_B (t'-t_1))}dt' = - 2 i \gamma T_B
\end{equation}
for the HFP normalization. This already holds to good approximation for finite temperatures so long as $\hbar \omega_\gamma \ll k_B T_B$ (in the underdamped regime) if one assumes that $f$ does not vary on timescales much shorter than $1/\omega_\gamma$, which is the case for the system's correlators. Then, as a distribution acting on this class of functions
\begin{equation}
    \fp \frac{i\gamma \pi T_B^2}{\sinh^2(\pi T_B (t'-t_1))} \overset{T_B/\omega_\gamma \to\infty}{\longrightarrow} -2i\gamma T_B \delta(t'-t_1)\,,
\end{equation}
which agrees with \cref{eq:trivial_order_classic_limit} and is exactly the white-noise memory kernel found in the Langevin equation of classical Brownian motion \cite{weiss_quantum_1999,kamenev_field_2011} (see also \cref{eq:Langevin}).

%% file: sections/Separation_of_Scales.tex
\section{\label{sec:separation of scales}Separation of scales}
In a structured environment with features on energy scales comparable to those of the system, there is no way around explicitly resolving those dynamics by adapting the time discretization in a numerical solver to the smallest relevant scale. On the other hand, there is a class of environments with structure on a characteristic scale $\omega_c$ much larger than the system's scales. For such environments, the relevant feature of the effective bath spectrum is its behaviour for $\omega \ll \omega_c$, e.g. $\Gamma(\omega)\propto\omega$ and its generalization $\Gamma(\omega)\propto\sgn(\omega)\abs{\omega}^s$. Consider, e.g., phonons coupled to the conduction band of electrons with a characteristic scale on the order of the bandwidth $\omega_c \sim W$ or a low-energy bosonic collective mode, such as the Higgs mode in superconductors, coupled to the quasiparticle band. In these settings, one can easily find $\omega_c/\omega_0 > 100$. Adapting the discretization scale to $\omega_c$ is catastrophic in such a case, as it leads to an $\mathcal{O}\left((\frac{\omega_c}{\omega_0})^3\right)$ fold increase in CPU time compared to the isolated system to simulate the same physical time. Instead, one would naturally wish to use a $\omega_c/\omega_0 \to \infty$ limit. However, we have seen before that in this limit, the momentum variance, and therefore the acceleration of the symmetric correlator, diverges. The next best thing is to derive a time-stepping scheme that retains the same discretization $\delta t \sim 1/\omega_0$ as one would use for the system alone. Still, it captures the $\omega_c$-dependent aspects of the dynamics with minimal computational cost. We will see in section \ref{sec:Renormalization} that in the case of an ohmic environment, this also enables renormalizing the momentum divergence and taking the $\omega_c \to \infty$ limit after all.\\

The ansatz we use is inspired by \eqref{eq:peel-off}, arising directly from interpreting the embedding self-energies as Hadamard finite-part distributions. The critical point in the convolutions $(\Sigma_ \emb \cdot G) (t_1,t_2)= \int_{t_0}^\infty dt' \Sigma_ \emb(t_1,t')G(t',t_2)$ is $t' = t_1$, where we probe the UV of $\Sigma_ \emb$. Let us expand the correlators around that critical point,
\begin{widetext}
\begin{equation}
    G(t',t_2) = G(t_1,t_2) + \partial_{t_1} G(t_1,t_2) (t' - t_1) + R_2(t_1,t',t_2)
\end{equation}
\begin{equation}
     \text{with}\qquad R_2(t_1,t',t_2) := G(t',t_2) - G(t_1,t_2) - \partial_{t_1} G(t_1,t_2) (t'-t_1)\,.
 \end{equation}
 \end{widetext}
Then, we can use the linearity of the integral to compute the convolution term by term. In the convolution $\Sigma^\ret\cdot G^S$, we find the familiar expression
\begin{equation}
\begin{split}
    \int_{t_0}^{t_1} \Sigma^A_\emb(t_1,t') dt' &= -\frac{2 \gamma \omega_c}{\pi \left(1 + \frac{1}{(t_1 \omega_c)^2}\right)}\\
    &\xrightarrow{t_1-t_0 \gg \omega_c\inv} - \frac{2\gamma \omega_c}{\pi}
\end{split} 
\end{equation}
for the first term, which is the same mass-renormalization we found in the equilibrium theory. Moreover, we also find
\begin{align}
    &\int_{t_0}^{t_1} \Sigma^A_\emb(t_1,t') (t'-t_1)dt' \nonumber \\
    &= - \frac{2\gamma}{\pi} \left(\frac{t_1 \omega_c}{1 + (t_1 \omega_c)^2} + \arctan(t_1 \omega_c)\right)\\
    &\xrightarrow{t_1-t_0 \gg \omega_c\inv} \gamma \nonumber \,,
\end{align} 
for the second term, providing the linear damping. Lastly, the remainder-term scales like
\begin{equation}
   \Sigma^\ret_\emb \cdot R_2 \xrightarrow{t_1 - t_0 \gg \frac{1}{\gamma}} \mathcal{O}\left(\frac{\gamma}{\omega_c}\log\left(\frac{\omega_c}{\gamma}\right)\right)
\end{equation}
such that we find for $t_1 \gg \gamma\inv$ and $\gamma \ll \omega_c$ the time-local approximation
\begin{equation}
    \left(\Sigma^\ret_\emb \cdot G^S\right) \approx - \frac{2\gamma \omega_c}{\pi} G^S + \gamma \partial_{t_1} G^S\,.
\end{equation}
The mass renormalization is canceled by the counterterm that we added to the bare resonance frequency. The convolution $\Sigma^\ret\cdot G^\ret$ in the antisymmetric correlator's KBE is treated analogously. The only difference is that the integral runs from $t_2$ instead of $t_0$ to $t_1$, hence the limits are replaced by $\abs{t_1-t_2}$ and the first-order term is replaced by $\gamma\, \sgn(t_1 - t_2)$.\\

Lastly, we consider $\Sigma^S\cdot G^\adv$. For later convenience, we define
 \begin{align}
     P(t_1,t_2) &:= \int_{t_0}^{t_2} \Sigma^S_ \emb(t_1,t') dt'\,,\\
      Q(t_1,t_2) &:= \int_{t_0}^{t_2} \Sigma^S_ \emb(t_1,t') (t'-t_1) dt'\,.
 \end{align}
which are independent of the system and its current state and can be solved analytically (see Appendix \ref{sec:zero_and_one}). The convolution is thus partitioned to read
\begin{widetext}
\begin{equation}\label{eq:memory_term_interpretation}
     \begin{split}
         \int_{t_0}^{t_2} \Sigma^S_ \emb(t_1,t') G^A(t',t_2) \,dt' &= \int_{t_0}^{t_2} \Sigma^S_ \emb(t_1,t') R^A_2(t_1,t',t_2) \,dt' + G^A(t_1,t_2)\, P(t_1,t_2) + \partial_{t_1} G^A(t_1,t_2)\, Q(t_1,t_2)
     \end{split}\, .
 \end{equation}
\end{widetext}
Because the symmetric embedding self-energy (aka. the memory kernel) does not decay over $1/\omega_c$ but $1/T_B$, it introduces non-markovianity and is not negligible even for large $\omega_c$. However, the quadratic singularity of the memory kernel for $\omega_c \to \infty$ is regularized by subtraction of the zeroth and first-order expansion terms, since as $t'\to t_1$, $R_2 \sim (t'-t_1)^2$ lifts $\Sigma^S \sim 1/(t'-t_1)^2$. The value at $t'=t_1$ is given by l'Hôpital's rule,
\begin{equation}\label{eq:memory_term_lhopital}
     \frac{i\gamma\pi T_B^2 R_2(t_1,t',t_2)}{\sinh^2(\pi T_B (t'-t_1))} \overset{t'\to t_1}{\longrightarrow} \frac{i\gamma}{2\pi}\partial_{t_1}^2 G^A(t_1,t_2)\,.
 \end{equation}
 Although $\partial_t^2G^A$ has a discontinuity
 % \footnote{A proper discontinuity only develops in the $\omega_c \to \infty$ limit, but on the coarse $\omega_0$-timescale $\delta_{t_1}^2 G^A(t_1,t_2)$ behaves the same as \eqref{eq:discontinuity_of_G_A} with a jump from approximately $-\gamma$ at $t_1 = t_2-\Delta t$, to $0$ at $t_1=t_2$, to approximately $+\gamma$ at $t_1=t_2+\Delta t$ so long as $\omega_c \gg \omega_0$. In general using the $\omega_c \to \infty$ limit of $\Sigma^A_ \emb$ is only justified because the resulting $G^A_{\infty}$ cannot be distinguished from the finite sequence element, $G^A_{\omega_c}$, on time-scales much coarser than $\sim \omega_c\inv$. Importantly, a finite discontinuity in $\partial_t^2G^A$ does not obstruct the well-definedness of the (one-sided) HFP (in the $\omega_c \to \infty$ limit), for which (one-sided) Hölder continuity of $\partial_t G_A$ at $t'=t_1$ suffices \cite{monegato_definitions_2009}. Since $\partial_t^2G^A(t',t_1)$, is bounded around $t'=t_1$, $\partial_t G_A(t',t_1)$ is even Lipschitz continuous.} 
 at $t_1=t_2$, the causal structure of the KBEs ensures a unique value, since the singularity at $t'=t_1$ where l'Hôpital needs to be used is approached only from the past, such that for $t_2=t_1$ the limit in \eqref{eq:memory_term_lhopital} is $t'\to t_1^-$, where we know $\partial_t^2G^A(t^-,t) = -\gamma$. Thus, the remainder term in \eqref{eq:memory_term_interpretation} converges for $\omega_c \to \infty$ also on the diagonal.\\
 We still expect to find a term responsible for the logarithmic divergence of $\partial_{t_1}^2G^S(t_1,t_2)\big\vert_{t_1=t_2}$. On the diagonal, the antisymmetry of $G^A$ ensures that the zeroth-order term vanishes, and by the CCR $\partial_{t_1}G^A(t_1,t_2) = -1$. Then, by exclusion of both other terms, the logarithmic $\omega_c$-scaling must stem from the Q-term. Indeed, we find
 \begin{widetext}
     \begin{equation}\label{eq:one_term}
    Q(t,t) =  \left(\frac{i\gamma}{2\pi}\log(\frac{1-\cos(\frac{2\pi T}{\omega_c})}{\cosh(2\pi T (t-t_0)) - \cos(\frac{2\pi T}{\omega_c})}) - i\gamma T \frac{2\pi T (t-t_0)\sinh(2\pi T (t-t_0))}{ \cos(\frac{2\pi T}{\omega_c}) - \cosh(2\pi T (t-t_0))}\right)
\end{equation}
 \end{widetext}
 which converges exponentially fast in $2\pi T (t-t_0)\to\infty $ to a constant logarithmically divergent jump/kick,
\begin{equation}\label{eq:jump}
    - \frac{i\gamma}{2\pi}\log(4\sin^2\left(\frac{\pi T}{\omega_c}\right)) \sim \frac{i\gamma}{\pi} \log(\frac{\omega_c}{2\pi T})\,.
\end{equation}
We have therefore successfully isolated the fast, $\omega_c$-dependent first-order term, $Q$, from the slow history dependence in the remainder term $R_2$.\\
Extending the subtraction \eqref{eq:memory_term_interpretation} to the entire time interval $(t_0,t_2)$ is practical from an implementation point of view. However, \eqref{eq:jump} suggests a scaling behavior that is consistent only with the asymptotics of the momentum variance (Section \ref{subsec:uncertainty_relations}) for $T>\omega_\gamma$. The logarithmic divergence occurs because the integrand scales like $1/(t'-t)$ for $t' \to t$ ($t=t_1=t_2$ the time on the diagonal). Importantly, it does so only over a finite timescale which is the shortest of the three scales $\{\omega\inv_\gamma, \gamma\inv ,T\inv \}$. Either the $1/(t'-t)$ scaling is lost first because $(t'-t)T\approx 1$ and the memory kernel is not well approximated anymore by $\sim 1/(t'-t)^2$, or the approximation that the propagator scales linearly $G^A(t',t) \sim (t'-t)$ breaks down, either for its oscillation when $(t'-t)\omega_\gamma\approx 1$ or for its decay when $(t'-t)\gamma \approx 1$. Hence, we recover for $T,\gamma \ll \omega_\gamma$, the scaling $\log(\omega_c/\omega_\gamma)$ and for $\gamma,\omega_\gamma \ll T$, the scaling $\log(\omega_c/T)$ as well as close to critical damping for $\omega_\gamma, T \ll \gamma$ the scaling $\log(\omega_c/\gamma)$. For our implementation this means, if the linear scaling is restricted by $G^A$ instead of $\Sigma^S_ \emb$, the subtraction of the linear term in the regulated integral (i.e. the first term in \eqref{eq:memory_term_interpretation}) does not cancel the linear scaling regime entirely, but restricts it to the interval between the fastest scale, $\omega\inv_\gamma$ or $\gamma\inv$, and $T\inv$. Hence, the regulated integral may acquire a contribution scaling as $\log(T/\omega_\gamma)$ or $\log(T/\gamma)$, thereby restoring the correct overall scaling.\\

Taking a step back, we can analyze the origin of the logarithmic jump in \eqref{eq:memory_term_interpretation}. It occurs because, at equal times: (a) the zeroth-order term vanishes; and (b) the first-order term is finite. However, (a) is always given by symmetry, and (b) is always given by the canonical commutation relation. This implies that the logarithmic $\omega_c$-scaling of the equal-time acceleration  $\partial_{t_1}^2G^S(t_1,t_2)\big\vert_{t_1=t_2}$ is a general feature of bosonic systems coupled to an ohmic environment. In a steady state, this automatically translates to the momentum variance $\expval{\pi(t)\pi(t)}$.\\
Both $\Sigma_ \emb^S$ and $G^A$ decay exponentially over long times, such that only in the region where $\abs{t'-t_2} \gamma \lesssim 1$ and $\abs{t'-t_1}T \lesssim 1$ the memory term in \eqref{eq:memory_term_interpretation} gets a meaningful contribution. This can be exploited to introduce a finite memory, beyond which the exponentially suppressed past is neglected, thereby limiting the size of the matrices that need to be multiplied. As a consequence, for times larger than the memory cutoff, the cost of each step grows only linearly; the time-stepping algorithm reduces to quadratic scaling, where the cost of each step depends linearly on the memory depth. Because of this, long-time evolutions are less computationally expensive at higher temperatures, although for very high temperatures, one might use the classical limit (\ref{subsec:classical_limit}) directly.\\

%% file: sections/Implementation.tex
\section{Numerical implementation}
\begin{figure}[t]
    \centering
    \includegraphics[width=\linewidth]{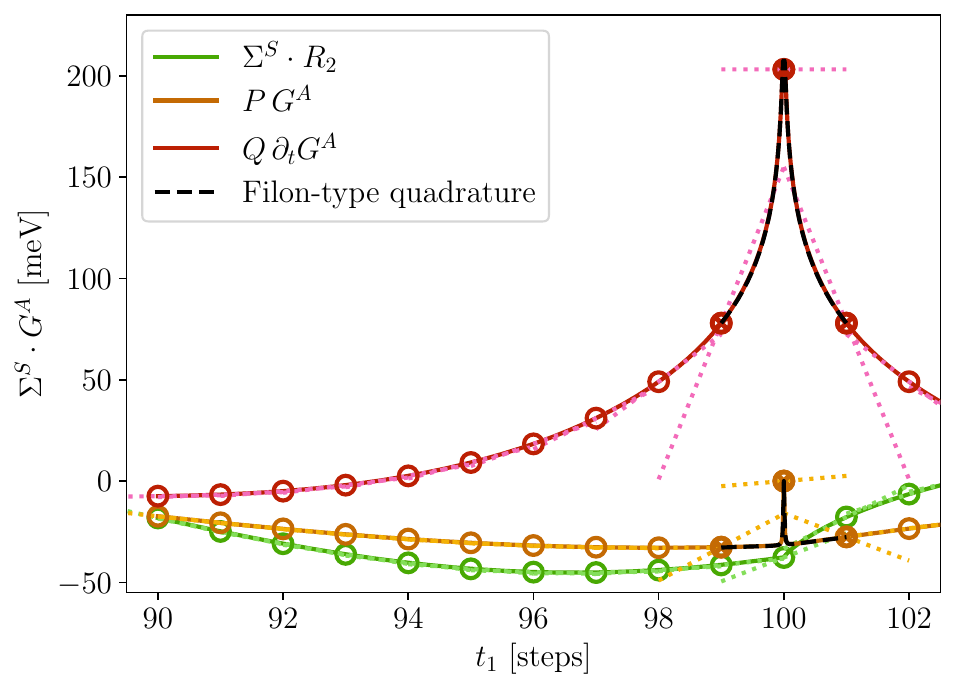}
    \caption{The terms of \eqref{eq:memory_term_interpretation} which constitute the nontrivial part of the $f$ term in \eqref{eq:toy_model} are shown as a function of $t_1$ for a fixed value of $t_2=100\, \Delta t$. The linear local approximations, implicit in the Verlet integration, are shown as dotted lines. X-markers highlight the points around $t_1=t_2$, where the linear approximation breaks down. Lastly, a black dashed line shows the Filon quadrature interpolation. The model parameters are $\omega_\gamma = 196$ meV, $\gamma = 200$ meV, $T = 26$ meV and $\omega_c = 100/\Delta t$, where $\Delta t = \frac{1}{30}\frac{2\pi}{\omega_0}$}
    \label{fig:implementation_vs_exact}
\end{figure}
So far, we have taken care of the separation of scales in the convolutions $\Sigma\cdot G$. This is the crucial step that makes the KBEs amenable to solvers using $\omega_0\inv$-scale discretization. Conceptually, the key principle behind such an implementation is an adiabatic approximation of the system compared to the environment. For simplicity, let us consider the KBE of $G^S(t_1,t_2)$ formally as an ODE in $t_1$ and hold $t_2$ fixed (i.e. use the method of lines),
\begin{equation}\label{eq:toy_model}
    \Ddot{y}(t_1) = f(y,\dot y,t_1) \,.
\end{equation}
The term we have regulated so far is part of the force $f$ and therefore must be integrated. However, $f$ is sharply peaked at $t_1=t_2$ due to the $\omega_c$-dependent terms, $P$ and particularly $Q$. We isolate the fast terms,
\begin{equation}\label{eq:separated_ODE}
    \Ddot{y}(t_1) = s(t_1) + \tilde f(y,\dot y,t_1)
\end{equation}
where
\begin{equation}
    s(t_1) = G^A(t_1,t_2)\,P(t_1,t_2) + \partial_{t_1}G^A(t_1,t_2)\,Q(t_1,t_2)
\end{equation}\\
and $\tilde f = f - s$ is varying slowly enough to be resolved by the system time-scale (cf. \ref{fig:implementation_vs_exact}). 
\begin{figure}[b]
    \centering
    \includegraphics[width=\linewidth]{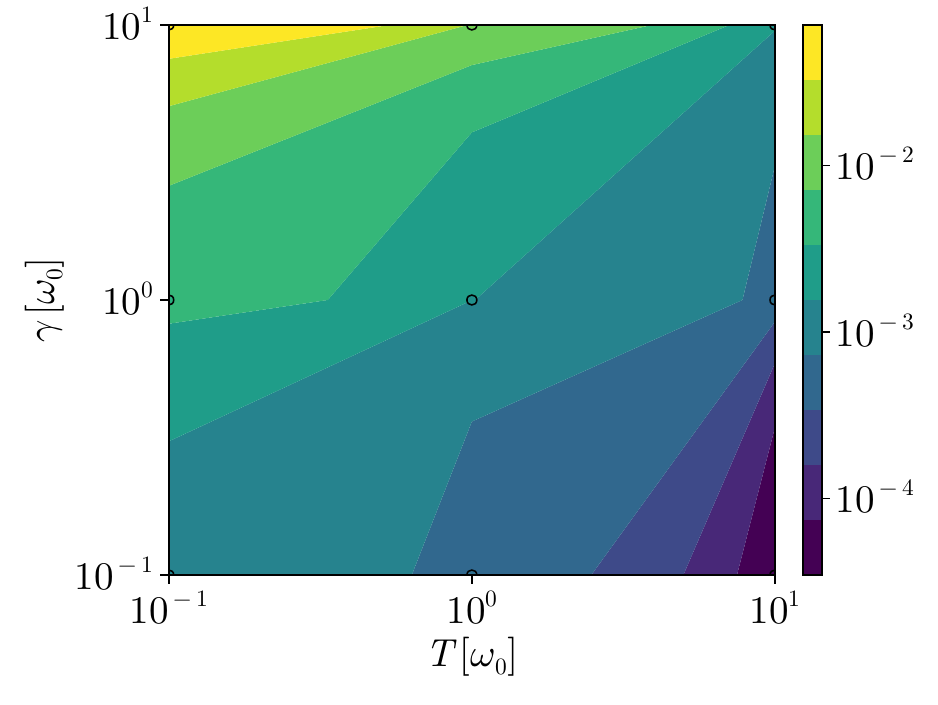}
    \caption{Thermalization benchmark of the numerical implementation across the temperature-damping parameter landscape. The relative error of the long-time asymptote $\lim_{t\to\infty}G^S(t,t)$ of the implementation with an environment at temperature $T$ compared to the exact thermal equilibrium variance $\expval{\varphi^2}_T$ obtained in the Matsubara formalism is shown. All axes use logarithmic scaling. The time step $\Delta t$ is $\frac{2\pi}{100}$ of the fastest system scale ($\omega_0$, $T$, or $\gamma$) and $\omega_c = 10^5 \omega_0$.}\label{fig:relative_errors}
\end{figure}
The ODE can now be viewed as having a locally stiff component, $s(t_1)$, and the integrator should be adapted accordingly. Since we are dealing with the linear damping case, a variant of exponential time differencing (ETD) takes full advantage of this linearity, but other integrators, e.g., linear damping-adapted Verlet integration, are suitable as well. The exact "position-only" time step is given by the formula
\begin{equation}
    y_{n+1} = a\,y_n - b\,y_{n-1} + \int_{-\Delta t}^{\Delta t} h(\tau) f(t_n + \tau) d\tau 
\end{equation}
where for Verlet integration we have $a=2$, $b=1$ and $h(\tau) = (\Delta t - \abs{\tau})$ and for the ETD step we subtract the terms linear in $G^S$ from $f$ and find (see appendix \ref{sec:ETD} for more details) 
\begin{equation}
    a = 2\cos(\omega_\gamma \Delta t) e^{-\frac{\gamma}{2}\Delta t}\,,\qquad b = e^{-\gamma \Delta t}
\end{equation} and 
\begin{equation}
    h(\tau) = e^{-\frac{\gamma}{2}(\Delta t - \tau )} \frac{1}{\omega_\gamma}\sin\left(\omega_\gamma (\Delta t - \abs{\tau})\right)\,.
\end{equation}
Now we can exploit the fact that the stiff term factorizes into a slow $G^A$ part and the $\omega_c$-dependent parts by using a Filon-type quadrature approximation for the integral over $s(t_1)$. We use the (quadratic) Lagrange interpolating polynomials $l_i$ on the three nodes $i\in -1,0,1$ to construct the weights
\begin{equation}\label{eq:weights}
    w^Q_i(t_1,t_2) := \int_{-\Delta t}^{\Delta t}d\tau h(\tau) Q(t_1 + \tau,t_2) l_i(\tau)
\end{equation}
which allows us to approximate
\begin{widetext}
    \begin{equation}\label{eq:Filon_Q}
    \int_{-\Delta t}^{\Delta t} h(\tau) \partial_t G^A(t_1 + \tau,t_2)Q(t_1 + \tau,t_2) d\tau \approx  \sum_{j=-1}^1 \partial_t G^A(t_1+j\Delta t,t_2)\,w^Q_{j}(t_1,t_2)\,.
\end{equation}
\end{widetext}
Because we use the Filon-type quadrature, the error is of the order $\sup_{\tau \in (-1,1)}\partial_t^4 G^A(t_1+\tau \Delta t,t_2) \Delta t^3$. Importantly, it does not rely on a slow variation of $Q(t_1,t_2)$. Exactly on the diagonal, i.e., $t_1=t_2$, we face the problem that $G^A$ has a discontinuity in its second derivative. In this case, we must divide the interval into the two sub-intervals $[-1,0]$ and $[0,1]$, and use linear interpolation on each. Doing this, we lose one order of $\Delta t$ accuracy, but gain the advantage that on the sub-intervals the derivative prefactors of the error terms are bounded. We can proceed analogously with the $P$-term; however, for $\omega_c \gg \omega_0$, the value of $G^A(t_1,t_2)P(t_1,t_2)$ on the diagonal converges to the constant $-i\gamma/\pi$, obtained using l'Hôpital's rule in the $\omega_c \to \infty$ limit. Since this renders the $P$-term smooth (the peak to zero in fig. \ref{fig:implementation_vs_exact} disappears), it may be treated as slowly varying and may be moved from $s$ to $\tilde f$ in \eqref{eq:separated_ODE}.\\
Notice that in a quadratic interpolation of $G^A(t_n + \tau, t_2)$ over the interval $[-\Delta t, \Delta t]$ we use the value $G^A(t_{n+1},t_2)$, "in the future" of the current state of $G^S$. In the non-interacting theory, this is possible because $G^A$ can be solved independently of $ G^S$. However, in an interacting theory where the KBEs are coupled, one must use the constant interpolation at $G^A(t_n,t_2)$. To improve accuracy, this result can be used as the predictor in a predictor-corrector step using quadratic interpolation.\\

Consequently, in the step $n \to (n+1)$, where $t_1 = t_n$, we use Filon-quadrature for the stiff term on the three lines $t_2\in \{t_{n-1},t_n,t_{n+1}\}$ (highlighted by the orange box in Figure \ref{fig:time_stepping}) while for all other $t_2 < t_{n-1}$ no overlap with the time diagonal occurs in the integration along $t_1$, such that the UV-dynamics is not resolved. There, the $P$ and $Q$ terms need not be separated from the remainder, and standard quadrature methods can be used. The weights for the Filon-quadrature can be computed with an adaptive quadrature routine. Importantly, after the memory time-scale $\sim T\inv$, both $P$ and $Q$ are translation invariant (see Appendix \ref{sec:zero_and_one}). Therefore, once the initial state is forgotten, the weights do not change and need not be recomputed.\\

\begin{figure}
    \centering
    \includegraphics[width=0.99\linewidth]{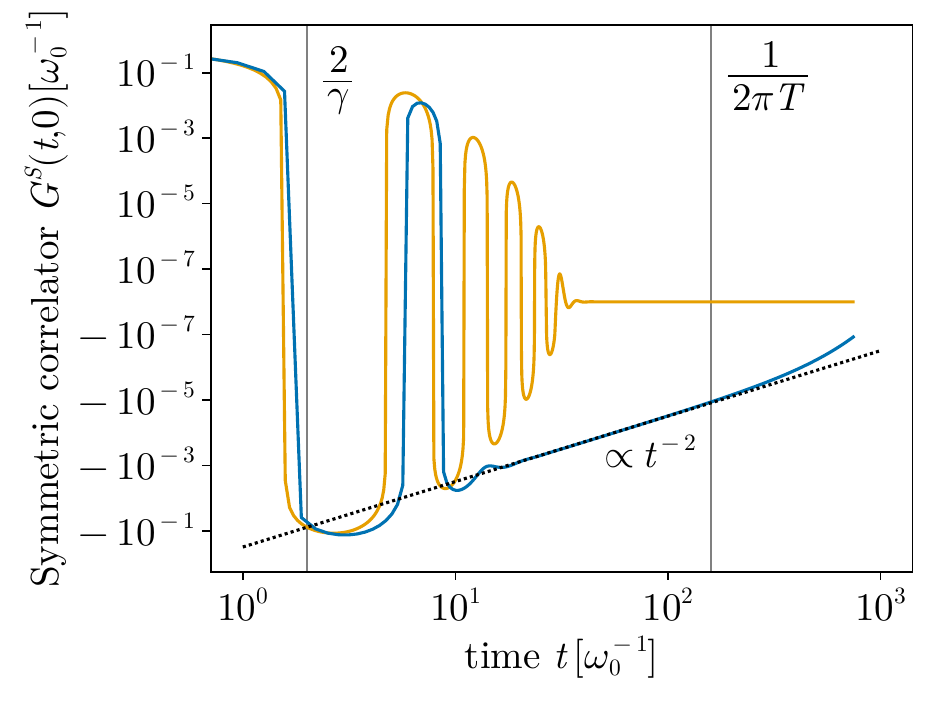}
    \caption{Decay behaviour of the symmetric correlator in the ultra-cold regime ($T<\gamma$). The KBE solution (blue) is compared against the solution of the Lindblad equation (yellow). Whereas the Lindblad solution shows immediate exponential decay, the KBE solution enters a regime of quadratic decay (black dotted line) on the time-scale between $2/\gamma$ and the first Matsubara frequency, $1/2\pi T$, indicated by vertical lines. Here, $\gamma = 1$ and $T = 0.001$.}
    \label{fig:decay}
\end{figure}

Using this implementation, the correlators indeed evolve from arbitrary initial conditions to the correct thermal-state correlators. We use the relative error of the amplitude variance $\expval{\varphi(t)^2} = i G^S(t,t)$ compared to its thermal expectation value $\expval{\varphi^2}_T$ as a proxy to determine the thermalization quality of the whole correlator $G^S$. A scan across the different parameter regimes is presented in Figure \ref{fig:relative_errors}. Clearly, the algorithm's accuracy decreases markedly in the regime of both very large damping and very low temperature. The source of the error in that regime is the term $\Sigma^S\cdot R_2$, which becomes highly localized around the diagonal. Because our focus, for now, lies on the underdamped regime, we do not add additional complexity to the time-stepping algorithm to make it suitable for arbitrarily overdamped configurations.\\
Nonetheless, the underdamped part of the non-Markovian regime with $T\ll \gamma \lesssim \omega_0$ is fully accessible (see fig. \ref{fig:decay}). In this regime, the symmetric correlator shows an algebraic $t^{-2}$ decay starting from the damping scale $\frac{2}{\gamma}$ and extending up to the temperature scale $\frac{1}{2\pi T}$, where exponential decay sets in \cite{jung_long-time_1985}. This is in contrast to the solution of the Lindblad formulation of the DQHO, which cannot capture this behaviour because it relies on the Born-Markov approximation. Using the Markov approximation for cold temperatures, one systematically underestimates long-time correlations and violates both the FDT and the KMS condition \cite{talkner_failure_1986}.

\subsection{Initial state preparation}
In the Keldysh formalism, the initial state can be encoded in the imaginary appendage $[t_0,t_0-i\beta)$ of the closed time contour. It is in principle possible to generate an arbitrary initial state via a suitable choice of the Hamiltonian $H^M$ on this contour \cite{stefanucci_nonequilibrium_2013}. Most relevant to experiments is the thermal equilibrium state generated by the simplest choice, $H^M=H(t_0)$. The right and left components, which are correlators between fields with real- and imaginary-time arguments, insert the initial-state correlation/memory into the KBE. A more detailed discussion of this can be found in \ref{sec:compact_past}.\\
In addition to correlations between the branches, the initial state determines the initial conditions for the KBEs, namely the expectation values of the respective correlators in that state. The KBEs (\ref{eq:antisymkbe}-\ref{eq:rightkbe}) are second-order differential equations. Therefore, we require two initial conditions each. The formulation in terms of symmetric and antisymmetric components makes determining these particularly simple. For the symmetric component, we find
\begin{equation}
    G^S(t_0,t_0) = -i\expval{\varphi^2}_\text{eq}\quad\text{and}\quad \partial_{t_1}G^S(t_0,t_0) =0\,,
\end{equation}
where the latter follows by symmetry and time translation invariance of the equilibrium state. For the antisymmetric component, the symmetry and canonical commutation relations provide
\begin{equation}
    G^A(t_0,t_0) = 0 \quad \text{and}\quad \partial_{t_1}G^A(t_0,t_0) = -1\,.
\end{equation}
Lastly, the right component's initial conditions are
\begin{align}
    G^\urcorner(t_0,\tau) &= G^M(\tau) \quad \text{and}\\
    \partial_{t_1}G^\urcorner(t_0,\tau) &= - \int_\omega \omega\, e^{\omega \tau} n_B(\omega) A(\omega)\label{eq:rightcomponent_initial_velocity}\,,
\end{align}
both of which follow from the identity $G^\urcorner(t,\tau) = G^<(t,t_0-i\tau)$ and using the thermal initial state, in particular the FDT for \eqref{eq:rightcomponent_initial_velocity}. Except for the antisymmetric component's initial conditions, which are fixed by symmetry and the CCR alone, all other initial conditions depend on the particular initial state.

\subsection{Coarse-graining}\label{sec:Renormalization}
\begin{figure}
    \centering
    \includegraphics[width=0.99\linewidth]{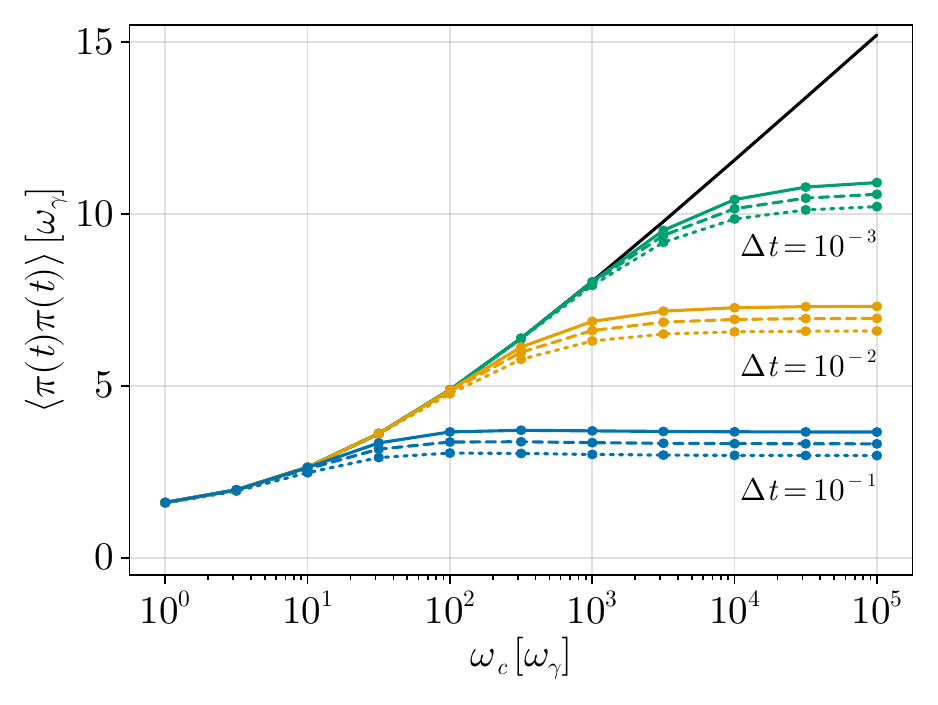}
    \caption{Log-linear plot of momentum variance $\expval{\pi^2}$ against the environment scale $\omega_c$. The solid black line shows the exact value. In contrast, the colored scatter-lines show different approximations on a grid with discretizations given by the value of $\Delta t$ below the respective group of graphs. For each discretization, we display $\mathcal{O}(\Delta t^2)$ (dotted) and $\mathcal{O}(\Delta t^4)$ (dashed) finite difference approximations of the second derivative, as well as the second derivative of the Whittaker-Shannon interpolation $\tilde G^S$ (full). The remaining model parameters used for the plot are $\gamma = 5$ and $T = 0.5$ in units of $\omega_\gamma$ (in the underdamped regime).}
    \label{fig:renormalization}
\end{figure}
Our implementation computes only the values of functions that converge in the limit $\omega_c \to \infty$, e.g., $G^S$, but not $\partial_{t_1}^2G^S$. Assuming these limiting functions are well-defined at every point (this excludes singularities and pathological functions like $\sin(\frac{1}{t})$), their derivatives are guaranteed to be locally integrable. In our case, we find $\partial_{t_1}^2G^S$ to diverge logarithmically at $t_1 = t_2$. However, this implies that in the time step
\begin{equation}
    y_{n+1} = a\, y_n - b\, y_{n-1} + \int_{-\Delta t}^{\Delta t}d\tau\, h(\tau) y''(t_n + \tau)\,,
\end{equation}
to evolve $G^S$ across the diagonal, we can take the limit $\omega_c \to \infty$ since the singularity that appears in the integrand is logarithmic and therefore integrable. In our particular implementation, we have already isolated the $\omega_c$-dependence into the $Q$-term, which is integrated against $h(\tau)$ (that is also integrable) to compute the Filon-quadrature weights. From the above reasoning, it follows that these weights converge in the limit $\omega_c \to \infty$.\\
This coarse-graining behaviour can also be observed from a different perspective. Let us assume we had used a perfect integrator such that $G_{nm} = G(t_n,t_m)$ (where, in a slight abuse of notation, both the continuous function and our version living on the grid are denoted $G$). The best possible interpolation is given by the  Whittaker–Shannon formula, 
\begin{equation}
    \tilde G(t_1,t_2)= \sum_{n,m} G_{nm}\, l_n(t_1) l_m(t_2)
\end{equation}
with the sinc-interpolant $l_n(t_1) = \sinc\left(\pi\frac{t_1 - n\Delta t}{\Delta t}\right)$.
The spectrum of this interpolation is the continuous spectrum restricted to the second Nyquist zone.
\begin{equation}
    \tilde G(\bar t,\omega) = G(\bar t,\omega)\; \chi\left(-\frac{2\pi}{\Delta t}\leq \omega \leq \frac{2\pi}{\Delta t}\right)\,.
\end{equation}
What it means to use an $\omega_0$-scale discretization is that the resonance peak, or more generally all spectral structures inherent to the system, should lie within this region. However, since $\omega_c\gg \omega_0$, the exponential decay sets in far outside of it. We have therefore replaced the exponential regulator $e^{-\frac{\abs{\omega}}{\omega_c}}$ by a sharp cut-off at $\abs{\omega}=\frac{2\pi}{\Delta t}$, a much smaller scale. The Fourier transform of $\partial_{t_1}^2\tilde G^S$ is given by $-\omega^2 \tilde G(\omega)$, but since the value of $(\partial_{t_1}^2G^S)_{nn}$ depends on the cut-off, we could only recover the accurate value for a given $\omega_c$ if we chose $\omega_c\Delta t  \ll 1$. Instead, we find an acceleration that depends on the discretization scale. One may think of this as a renormalization procedure that averages the microscopic expression over the discretization scale. Finite-difference approximations to $\partial_{t_1}^2 G^S$ using the grid values $G^S_{nm}$ approach the interpolation derivative at infinite order, thus they present the same behaviour (see fig. \ref{fig:renormalization}). Although we can still obtain the unrenormalized expression by evaluating the KBE's RHS exactly on the diagonal, the resulting value is irrelevant to the dynamics on the coarse system time-scale.\\

%% file: sections/conclusion.tex
\section{\label{sec:Conclusion}Summary and Outlook}
We presented a new approach that enables efficient implementation of featureless environments within the Schwinger-Keldysh formalism, using the quantum harmonic oscillator in an Ohmic bath as a showcase. We attributed the well-known logarithmic divergence of the momentum uncertainty in the equilibrium state of the ohmically damped oscillator to a singularity in the Kadanoff-Baym equations (KBE), which is universal for ohmically damped bosonic systems. By identifying the symmetric component of the embedding self-energy as a nascent Hadamard finite part distribution, we were able to take the limit $\omega_c \to \infty$ analytically for all off-diagonal times. To resolve the remaining logarithmic singularity on the time-diagonal, we decomposed the convolution integral into fast, regulator-dependent local terms and slow, history-dependent contributions. Leveraging this scale separation, a time-stepping algorithm that integrates the local stiffness via Filon-type quadrature was developed. Crucially, it was demonstrated that resolving the dynamics on the coarse system time scale effectively renormalizes the locally integrable divergence, thereby permitting a rigorous transition to the ohmic wide-band limit.\\

Analogously, this regularization procedure is directly applicable to the standard wide-band limit embedding of fermionic systems. Here, the constant rate function,
\begin{equation}
    \Gamma(\omega) = \gamma
\end{equation}
results in a symmetric component
\begin{equation}
    \Sigma^S(t) \xrightarrow{\omega_c \to \infty} - \pv \frac{\gamma T}{\sinh(\pi T t)}\,,
\end{equation}
$\pv$ denoting the Cauchy principal value (also a HFP [see appendix \ref{sec:HFP}]). Because the fermionic anti-commutation relation dictates the fermionic advanced propagator's value on the diagonal, the convolution $\Sigma^S \cdot G^\adv$ leads to the same logarithmic divergence as in the bosonic case.\\

 Lastly, we give an outlook for more general bosonic rate functions,
 \begin{equation}
     \Gamma(\omega) = 2\gamma \omega \abs{\omega}^{s-1} e^{-\frac{\abs{\omega}}{\omega_c}}\,.
 \end{equation}
 The symmetric embedding self-energy has a singularity of order $s+1$. We use our interpretation as a nascent Hadamard finite part. The only divergence occurs on the diagonal where we need to subtract now $\floor{s+1}$ terms. For non-ohmic bath spectra, the smoothness properties of $G^A$ change. This is most transparent by considering the spectral function, which is the Fourier transform of the antisymmetric correlator
 \begin{equation}
     A_s(\omega) = \frac{2\gamma \omega \abs{\omega}^{s-1}}{(\omega^2-\omega_0^2)^2 + (\gamma \omega \abs{\omega}^{s-1})^2}
 \end{equation}
 The decay behaviour dictates how often $G^A(t)$ is differentiable, i.e. for a decay $\sim \abs{\omega}^{-n}$, $G^A(t)$ is at least in $C^{n-2}$, since $\omega^{n-2} A(\omega)$ absolutely converges. Interestingly, this behaviour does not depend monotonically on $s$; instead, we find
 \begin{equation}
     A_s(\omega) \xrightarrow{\omega \to \infty} \begin{cases}
         \abs{\omega}^{s-4}\, ,s\leq 2\\
         \abs{\omega}^{-s}\, , s\geq 2
     \end{cases}
 \end{equation}
 which implies a minimal decay at $s=2$ and stronger decay, and therefore smoother $G^A(t)$ for both $s<2$ and $s>2$.\\
 For sub-ohmic damping, $s<1$, there is no singularity in the KBE since $G^A(0)=0$ cancels the singularity in $\Sigma^S$ down to $t^{-s}$, which is integrable. So while for $\omega_c\to\infty$, $\Sigma^S$ is still singular, $\Sigma^S\cdot G^A$ converges.\\
 For super-ohmic damping, two sub-regimes can be distinguished. First for $1<s<2$, we still only have the $P$ and $Q$ terms with the $Q$ term now diverging rationally as $1/\abs{t}^{s-1}$. This divergence is integrable, as in the ohmic case, and is therefore renormalized. On the other hand, for $s\ge 2$, $Q$ has a non-integrable singularity. In this case, even the $\Delta t$-averaged/renormalized momentum variance does not converge for $\omega_c \to \infty$. Likewise, the variance of the amplitude does not converge. Therefore, taking the limit is not physically meaningful, and a finite value must be used.

%% file: sections/Acknowledgment.tex
\section*{\label{sec:Acknowledgment} Acknowledgments}
The idea for this paper arose from a problem that came up in a project to be published in collaboration with Michael Rübhausen and Dirk Manske. We would furthermore like to thank Henrik Müller-Groeling, Sida Tian, and Paulo Forni for fruitful discussions as well as Pietro Bonetti and Dirk Manske for useful comments on the manuscript.

%% file: sections/appendices/Caldeira_Leggett_model.tex
In this appendix, we consider the Caldeira-Leggett model as a possible environment and based on this show how the embedding self-energy arises. The Caldeira-Leggett model is generally given by the Hamiltonian
\begin{equation}
    H = \frac{P^2}{2M} + V(X) + \sum_i\left(\frac{p_i^2}{2m_i}+\frac{1}{2}m_i\omega_i^2 (x_i-\frac{C_i}{m_i\omega_i^2} X)^2\right)
\end{equation}
where $X,P$ denote the amplitude and momentum of the system oscillator and $x_i,p_i$ denote the amplitude and momentum of an infinite set of harmonic oscillators that form the environment. The coupling occurs in the “translation invariant” way, $(x_i-\frac{C_i}{m_i\omega_i^2} X)^2$ which gives rise to a much-discussed counterterm, $ \sum_i \frac{C_i^2}{2 m_i\omega_i^2}X^2$. In the language of a $0+1$ dimension QFT the Caldeira-Leggett model Lagrangian for a QHO system reads
\begin{equation}
    \mathcal{L}_{CL} = -\frac{1}{2}\varphi_0\left(\partial_t^2 + \omega_{0}^2\right)\varphi_0 - \sum_{i=1}^\infty c_i \varphi_0 \varphi_i - \frac{1}{2} \sum_{i=1}^\infty\varphi_i\left(\partial_t^2 + \omega_i^2\right)\varphi_i\,,
\end{equation}
where we have rescaled all the fields and the couplings $c_i$, to absorb the respective masses and defined the bare resonance frequency of the system $\omega_{0}^2 = \omega_{0,r}^2 + \sum_i \frac{c_i^2}{\omega_i^2}$ using the counterterm.
The Lagrangian is entirely quadratic/bilinear, and we can write it using the super-field $\varphi^T = (\varphi_0\,,\varphi_1\,\dots)$ in matrix form as
\begin{equation}
    \mathcal{L}_{CL} = -\frac{1}{2}\varphi^T G^{-1} \varphi\quad\text{with}\quad G^{-1} = \begin{pmatrix}
        g_0^{-1} & c_1 & c_2 & \dots \\
        c_1 & g_1^{-1} & 0 & \dots \\
        c_2 & 0 & g_2^{-1} & \dots \\
        \vdots & \vdots & \vdots & \ddots
    \end{pmatrix} \,.
\end{equation}
Here we use lower-case $g_i^{-1} = -\left(\partial_t^2+\omega_i^2\right)$ for the bare propagators and upper-case $G$ for the “dressed” one, since zero is already used as an index for the system.
Then, we can obtain the propagators via matrix inversion. In particular,
\begin{equation}
    G_{00} = \left(g_0^{-1} - \sum_{i=1}^\infty c_i^2 g_i\right)^{-1}
\end{equation}
and we can directly read off the embedding self-energy as
\begin{equation}
    \Sigma_ \emb(t,t') = \sum_i c_i^2 g_i(t,t')
\end{equation}
and therefore the rate-function in terms of the bath oscillator spectral functions
\begin{equation}
    \Gamma_ \emb(\omega) = \sum_i c_i^2 A_i(\omega) = \sum_i c_i^2 2\pi\, \sgn(\omega)\delta(\omega^2-\omega_i^2)
\end{equation}
One can then tune the infinite set of parameters $\omega_i,c_i$ to obtain essentially arbitrary functions $\Gamma(\omega)$. For the choice $c_i = \sqrt{\frac{\gamma}{\pi}}\omega_i$, the coupling strength scales linearly with the resonance frequency, and with a continuous set of $\omega_i$s one gets (using $\delta(\omega^2-\omega_i^2) = (\delta(\omega-\omega_i)+\delta(\omega+\omega_i))/2\abs{\omega_i}$)
\begin{equation}
    \Gamma_ \emb(\omega) = \gamma \int_0^\infty d\omega_i \frac{\omega_i^2}{\abs{\omega_i}}\sgn(\omega)\left[\delta(\omega-\omega_i) + \delta(\omega+\omega_i)\right)] = \gamma \omega
\end{equation}
which is the ohmic spectrum. Notice that we did not integrate out the bath. In a more general setting, instead of identifying directly the inverse propagator in the Lagrangian one can use the effective action formalism to obtain the 2-point functions, $\Gamma^{(2)}_{ij} := \delta^2\Gamma[\phi]/\delta \phi_i\delta \phi_j$. As long as the system-environment coupling remains bilinear, and the different bath oscillators are only coupled to one another via the system, no complications arise. In that case
\begin{equation}
    G_{00} = \left(g_0^{-1} - \Sigma_c - \Sigma_ \emb\right)
\end{equation}
with
\begin{equation}
    \Sigma_ \emb = \sum_i c_i^2 (g_i^{-1} - \Sigma_i)^{-1}
\end{equation}
where $\Sigma_i$ is the i-th bath mode's self-energy due to, e.g., anharmonicity or due to it being damped itself.

%% file: sections/appendices/finite_wc_mass_renorm.tex
\onecolumngrid
The renormalized “mass” $\bar\omega_0$ (in distinction to the resonance frequency, which is not the same for a finite imaginary part of the self-energy) is given by the root of
\begin{equation}
    \omega^2-\omega_0^2 - \Re \Sigma^\ret_ \emb(\omega)\,.
\end{equation} 
We can compute the self-energy's real part using the spectral representation and the rate-function \eqref{eq:regulated_rate_function} to be
\begin{equation}
    \begin{split}
        \Sigma^\ret_ \emb(\omega) &= \int_{-\infty}^\infty \frac{d\lambda}{2\pi}\frac{\Gamma(\lambda)}{\omega + i0^+ -\lambda} = \int_{-\infty}^\infty \frac{d\lambda}{2\pi} \left(\pv\frac{\Gamma(\lambda)}{\omega-\lambda} - i\pi \delta(\omega-\lambda) \Gamma(\lambda)\right)\\
        &= -\frac{2\gamma \omega_c}{\pi}+\frac{\gamma \omega}{\pi}\left(e^{-\omega/\omega_c} \text{Ei}(\omega/\omega_c)-e^{\omega/\omega_c} \text{Ei}(-\omega/\omega_c)\right) - i \gamma \omega e^{-\abs{\omega}/\omega_c}
    \end{split}
\end{equation}
with $\text{Ei}(z) = \int_{-\infty}^z \frac{e^{t}}{t}dt$. The second term is a correction that decays to zero for $\abs{\omega} \ll \omega_c$. For $\abs{\omega} \gg \omega_c$ it converges to $2\gamma \omega_c/\pi$ such that also the real part decays to zero.

%% file: sections/appendices/Fourier_trf.tex
\onecolumngrid
In this appendix, we present a detailed derivation of the time parametrized  embedding self-energies \eqref{eq:self-energy_A} and \eqref{eq:self-energy_S}, which is obtained via Fourier transformation of the frequency parametrized version. The symmetric one is known from the FDT and the rate-function to be 
\begin{equation}
    \Sigma^S(\omega) = -\frac{i}{2}\Gamma(\omega)\coth(\frac{\omega}{2T}) = -i\gamma \omega e^{-\abs{\omega}/\omega_c} \coth\left(\frac{\omega}{2T}\right)\,,
\end{equation}
as stated in \eqref{eq:Sigma_S_freq} of the main text. We find the Fourier transform
%\begin{widetext}
    \begin{equation}\label{eq:Keldysh_component_derivation}
    \begin{split}
        \Sigma_ \emb^S(t) &= -i\gamma \int_{-\infty}^\infty \frac{d\omega}{2\pi} e^{-i\omega t} \omega e^{-\abs{\omega}/\omega_c} \coth\left(\frac{\omega}{2T}\right) = \gamma \partial_t \int_{0}^\infty \frac{d\omega}{2\pi} \left(e^{-i\omega t} - e^{i\omega t}\right) e^{-\omega/\omega_c} \left(n_B(\omega) - n_B(-\omega) \right)\\
        &= 2\gamma \partial_t \int_{0}^\infty \frac{d\omega}{2\pi} \left(e^{-i\omega t} - e^{i\omega t}\right) e^{-\omega/\omega_c} \left( \sum_{n=1}^\infty e^{-\omega/T n} + \frac{1}{2} \right)\\
        &= 2 \gamma \sum_{n=1}^\infty \partial_t \int_{0}^\infty \frac{d\omega}{2\pi} \left(e^{-\omega(\frac{n}{T} + \frac{1}{\omega_c} +it)} - e^{-\omega(\frac{n}{T} + \frac{1}{\omega_c} - it)} \right) + \gamma \partial_t \int_{0}^\infty \frac{d\omega}{2\pi} \left(e^{-\omega( \frac{1}{\omega_c} +it)} - e^{-\omega( \frac{1}{\omega_c} - it)} \right)\\
        &=  \frac{\gamma}{\pi}\sum_{n=1}^\infty \partial_t  \left(\frac{1}{\frac{n}{T} + \frac{1}{\omega_c} +it} - \frac{1}{\frac{n}{T} + \frac{1}{\omega_c} - it} \right) + \frac{\gamma}{2\pi}\partial_t  \left(\frac{1}{ \frac{1}{\omega_c} +it} - \frac{1}{\frac{1}{\omega_c} - it} \right)\\
        &=  \frac{\gamma}{2\pi}\partial_t \left( 2\sum_{n=0}^\infty   \left(\frac{1}{\frac{n}{T} + \frac{1}{\omega_c} +it} - \frac{1}{\frac{n}{T} + \frac{1}{\omega_c} - it} \right) -   \left(\frac{1}{ \frac{1}{\omega_c} +it} - \frac{1}{\frac{1}{\omega_c} - it} \right)\right)\\
        &=  -i\frac{\gamma T^2}{\pi}  \left(\Psi^{(1)}\left(T(it + \frac{1}{\omega_c})\right) + \Psi^{(1)}\left(T(-it + \frac{1}{\omega_c})\right)\right) + i \frac{\gamma}{2\pi}  \left(\frac{1}{ (\frac{1}{\omega_c} +it)^2} + \frac{1}{(\frac{1}{\omega_c} - it)^2} \right)\\
        &=  -i\frac{\gamma T^2}{\pi}  \left(\Psi^{(1)}\left(1 + T(it + \frac{1}{\omega_c})\right) + \Psi^{(1)}\left(T(-it + \frac{1}{\omega_c})\right)\right) + i \frac{\gamma}{2\pi}  \left( \frac{1}{(\frac{1}{\omega_c} - it)^2 } - \frac{1}{(\frac{1}{\omega_c} + it)^2} \right)\\
        &=  -i\frac{\gamma T^2}{\pi}  \left(\Psi^{(1)}\left(1 + T(it + \frac{1}{\omega_c})\right) + \Psi^{(1)}\left(T(-it - \frac{1}{\omega_c})\right) + \sum_{n=0}^\infty \frac{\frac{4T}{\omega_c }\left(iTt - n\right)}{((iTt-n)^2-\frac{T^2}{\omega_c^2})^2}\right) + i \frac{\gamma}{2\pi}  \left( \frac{4 i t \frac{1}{\omega_c}}{(\frac{1}{\omega_c^2} + t^2)^2} \right)\\
        &=  -i\frac{\gamma }{\pi}  \frac{T^2\pi^2}{\sin^2\left(i\pi T t + \frac{T\pi}{\omega_c}\right)} -i\frac{\gamma T^2}{\pi} \sum_{n=0}^\infty \frac{\frac{4T}{\omega_c }\left(iTt - n\right)}{((iTt-n)^2-\frac{T^2}{\omega_c^2})^2} - \frac{2\gamma}{\pi}  \left( \frac{t\frac{1}{\omega_c}}{(\frac{1}{\omega_c^2} + t^2)^2} \right)\\
        &=  i\gamma \frac{T^2\pi}{\sinh^2\left(\pi T t - i\pi\frac{T}{\omega_c}\right)} -i\frac{\gamma}{\pi} \sum_{n=0}^\infty \frac{\frac{4}{\omega_c }\left(it - \frac{n}{T}\right)}{((it-\frac{n}{T})^2-\frac{1}{\omega_c^2})^2} - \frac{2\gamma}{\pi}  \left( \frac{t\frac{1}{\omega_c}}{(\frac{1}{\omega_c^2} + t^2)^2} \right)\\
        &=  i\gamma \pi T^2 \left(i\frac{2\sin\left(\frac{2\pi T}{\omega_c}\right)\sinh(2\pi t T)}{\left(\cos\left(\frac{2\pi T}{\omega_c}\right)-\cosh(2\pi t T)\right)^2} 
        + \frac{2\cos\left(\frac{2\pi T}{\omega_c}\right)\cosh\left(2\pi Tt\right) - 2}{\left(\cos\left(\frac{2\pi T}{\omega_c}\right)-\cosh(2\pi t T)\right)^2}\right)\\
        &\quad -i\frac{\gamma}{\pi} \sum_{n=0}^\infty \frac{\frac{4}{\omega_c }\left(it - \frac{n}{T}\right)}{((it-\frac{n}{T})^2-\frac{1}{\omega_c^2})^2} - \frac{2\gamma}{\pi}  \left( \frac{t\frac{1}{\omega_c}}{(\frac{1}{\omega_c^2} + t^2)^2} \right)\\
    \end{split}
\end{equation}
%\end{widetext}

where we used the geometric series to replace the Bose-Einstein distributions, going from line one to two, the trigamma function's series representation,
\begin{equation}\label{eq:Trigamma_series_rep}
    \Psi^{(1)}(z) = \sum_{n=0}^\infty \frac{1}{(z+n)^2}\,,
\end{equation}
going from line five to six, the trigamma's recurrence relation,
\begin{equation}
    \Psi^{(1)}(1+z) = \Psi^{(1)}(z) - \frac{1}{z^2}\,,
\end{equation}
going from line six to seven, and (using \eqref{eq:Trigamma_series_rep} again) the identity
%\begin{widetext}
    \begin{equation}
    \Psi^{(1)}(-iTt + \frac{T}{\omega_c}) - \Psi^{(1)}(-iTt - \frac{T}{\omega_c}) = \sum_{n=0}^\infty \frac{\frac{4T}{\omega_c }\left(iTt - n\right)}{((iTt-n)^2-\frac{T^2}{\omega_c^2})^2}\,,
\end{equation}
going from line seven to eight, as well as the reflection formula
\begin{equation}
    \Psi^{(1)}(1-z) + \Psi^{(1)}(z) = \frac{\pi^2}{\sin^2(\pi z)}\,,
\end{equation}
going from line eight to nine. All real terms in the result cancel (this is necessarily so, since the Fourier transformation of an imaginary-valued and even function is imaginary-valued and even too) since
\begin{equation}
    \frac{\gamma T^2}{\pi}\Im\left[\Psi^{(1)}\left(-iTt + \frac{T}{\omega_c}\right) - \Psi^{(1)}\left(-iTt - \frac{T}{\omega_c}\right)\right] = \gamma \pi T^2 \frac{2\sin\left(\frac{2\pi T}{\omega_c}\right)\sinh(2\pi t T)}{\left(\cos\left(\frac{2\pi T}{\omega_c}\right)-\cosh(2\pi t T)\right)^2} + \frac{2\gamma}{\pi}  \left( \frac{t\frac{1}{\omega_c}}{(\frac{1}{\omega_c^2} + t^2)^2} \right)\,.
\end{equation}
Consequently, the symmetric embedding self-energy of the exponentially regulated ohmic bath is exactly given by
\begin{equation}
    \Sigma^S_ \emb(t) = i\gamma \pi T^2 \frac{2\cos\left(\frac{2\pi T}{\omega_c}\right)\cosh\left(2\pi T(t_1-t_2)\right) - 2}{\left(\cos\left(\frac{2\pi T}{\omega_c}\right)-\cosh(2\pi T(t_1-t_2) )\right)^2} - i\frac{\gamma T^2}{\pi}\Re\left[\Psi^{(1)}\left(-iTt + \frac{T}{\omega_c}\right) - \Psi^{(1)}\left(-iTt - \frac{T}{\omega_c}\right)\right]\,.
\end{equation}
%\end{widetext}
The second term is abbreviated in the main text as $R_{\omega_c}$. It scales globally like $T/\omega_c$ for $\omega_c\to\infty$ and can be neglected for large $\omega_c$.\\

The anti-symmetric embedding self-energy is obtained more easily to be
\begin{equation}
    \begin{split}
        \Sigma^A_ \emb(t) &= -i \int_{-\infty}^\infty \frac{d\omega}{2\pi} \Gamma(\omega) e^{-i\omega t} = -2i\gamma \int_{-\infty}^\infty \frac{d\omega}{2\pi} \omega \,e^{-\abs{\omega}/\omega_c} e^{-i\omega t} \\
        & = 2\gamma \partial_t \int_{0}^\infty \frac{d\omega}{2\pi} \left(e^{-\omega(1/\omega_c + it)} + e^{-\omega(1/\omega_c - it)}\right) = \frac{\gamma}{\pi} \partial_t \left(\frac{1}{it + \omega\inv_c}+ \frac{1}{- it + \omega\inv_c}\right) = \frac{\gamma}{\pi} \partial_t \frac{2\omega\inv_c}{t^2 + \omega^{-2}_c}\\
        &= - 2\gamma \frac{2 t/\omega_c}{\pi(t^2 + \omega_c^{-2})^2}
    \end{split}
\end{equation}

The imaginary-time Matsubara component is obtained analogously. However, there are important differences with respect to the relevance of the different terms on the imaginary contour. First, we can analytically continue the spectral representation of the retarded component
\begin{equation}
    \Sigma^\ret(\omega) = \int_\lambda \frac{\Gamma(\lambda)}{\omega + i0^+ - \lambda} \quad\to\quad \Sigma(z) = \int_\lambda \frac{\Gamma(\lambda)}{z - \lambda}\quad\text{for}\, z\in\mathbb{C}
\end{equation}
Then the Matsubara component has the spectral form,
\begin{equation}
    \Sigma^M(i\omega_n) = \int_{-\infty}^\infty \frac{d\omega}{2\pi} \frac{\Gamma(\omega)}{i\omega_n-\omega} = -\int_0^\infty \frac{d\omega}{\pi} \frac{\omega\,\Gamma(\omega)}{\omega^2 + \omega_n^2}\,.
\end{equation}
The Fourier transformation can be written in the standard form
\begin{equation}
\begin{split}
    \Sigma^M(\tau) &= iT\sum_n e^{-i\omega_n\tau}\int_{0}^\infty \frac{d\omega}{\pi} \frac{\omega\,\Gamma(\omega)}{(i\omega_n - \omega)(i\omega_n + \omega)} = -i\int_0^\infty \frac{d\omega}{2\pi} \Gamma(\omega) \frac{\cosh(\omega(\beta/2-\abs{\tau}))}{\sinh(\omega \beta/2)}\,.
\end{split}
\end{equation}
Here and in the following, the absolute value should be understood as the absolute value on the reals, or simply as shorthand notation for the lengthy expression $f(\abs{z}) = \theta(\Re z) f(z) + \theta(-\Re z) f(-z)$. This distinction is relevant when analytically continuing the expression.
Inserting the explicit expression for the rate-function, one finds
\begin{equation}
\begin{split}
    \Sigma^M(\tau) = -\frac{i\gamma T^2}{\pi}\left[\Psi^{(1)}\left(1 + (T(\omega_c\inv - \abs{\tau})\right) + \Psi^{(1)}\left(T(\omega_c\inv + \abs{\tau})\right)\right] 
\end{split}
\end{equation}
The $\gtrless$ - components can be read off from the expression for $\tau\gtrless 0$ respectively. They are analytic, and it holds that
\begin{equation}
    \Sigma_M^\gtrless(z) = \Sigma^\gtrless(-iz)\,.
\end{equation}
From this one can also deduce $\Sigma^M(\tau) = \Sigma^T(-i\tau + \sgn(\tau)0^+)$ and $\Sigma^M(it + \sgn(t)0^+) = \Sigma^T(t)$.

%% file: sections/appendices/HFP_identities.tex
%\begin{widetext}
In this section, we show that the integrands we use in the main text are indeed a nascent CPV, that is
\begin{equation}\label{eq:PV_memory_kernel}
    \dashint_a^b \coth(\pi T(t'-t))f(t') dt' = \lim_{\varepsilon\to 0}\int_a^b \frac{\tanh(\pi T(t'-t))}{\tanh^2(\pi T(t'-t)) + \varepsilon^2}f(t') dt'
\end{equation}
and a nascent HFP,
\begin{equation}\label{eq:FP_memory_kernel}
    \ddashint_a^b\frac{f(t')}{\sinh^2(\pi T(t'-t))}dt' = \lim_{\varepsilon\to 0}\int_a^b \frac{\left(\sinh^2(\pi T(t'-t))(1-2\varepsilon^2) - \varepsilon^2\right)}{(\sinh^2(\pi T(t'-t)) + \varepsilon^2)^2}f(t') dt'
\end{equation}
for $f\in C^{1,1}$ and $a<t<b$.\\
Let us begin with the \textbf{CPV} \eqref{eq:PV_memory_kernel}. By definition, the LHS reads
\begin{equation}
    \begin{split}
        \dashint_a^b \coth(\pi T(t'-t))f(t') dt' &= f(t) \dashint_a^b \coth(\pi T(t'-t))dt' + \int_a^b\frac{R_1(t';t)}{\tanh(\pi T (t'-t))}\\
        &= \frac{f(t)}{\pi T} \log\abs{\frac{\sinh(\pi T(b-t))}{\sinh(\pi T(a-t))}} +\int_a^b\frac{R_1(t';t)}{\tanh(\pi T (t'-t))}
    \end{split}
\end{equation}
In the first step, we used Taylor expansion to isolate the orders needing explicit regularization (here only $n=0$) from the higher orders, which have regular integrands and for which the principal value coincides with the standard integral. This is the tactic most manipulations of (hyper-)singular integrals have in common if one assumes sufficiently smooth test functions. The RHS is
\begin{equation}
    \begin{split}
        &\lim_{\varepsilon\to 0}\int_a^b \frac{\tanh(\pi T(t'-t))}{\tanh^2(\pi T(t'-t)) + \varepsilon^2}f(t') dt'\\ 
        &= f(t) \lim_{\varepsilon\to 0}\int_a^b \frac{\tanh(\pi T(t'-t))}{\tanh^2(\pi T(t'-t)) + \varepsilon^2}dt' + \int_a^b\frac{R_1(t';t)}{\tanh(\pi T (t'-t))}\\
        &=\frac{f(t)}{\pi T} \lim_{\varepsilon\to 0}\frac{2\log(\cosh(\pi T(t'-t))) + \log(\varepsilon^2+\tanh^2(\pi T(t'-t)))}{2+2\varepsilon^2}\Big\vert_a^b + \int_a^b\frac{R_1(t';t)}{\tanh(\pi T (t'-t))}\\
        &= \frac{f(t)}{\pi T} \left(\log(\frac{\cosh(\pi T(b-t))}{\cosh(\pi T(a-t))}) + \frac{1}{2}\log(\frac{\tanh^2(\pi T(b-t))}{\tanh^2(\pi T(a-t))})\right) + \int_a^b\frac{R_1(t';t)}{\tanh(\pi T (t'-t))}\\
        &= \frac{f(t)}{\pi T} \log\abs{\frac{\sinh(\pi T(b-t))}{\sinh(\pi T(a-t))}} + \int_a^b\frac{R_1(t';t)}{\tanh(\pi T (t'-t))}
    \end{split}
\end{equation}
We continue with the \textbf{HFP} \eqref{eq:FP_memory_kernel} along the same lines and exploit the identity we just showed.
\begin{equation}\label{eq:FP_kernel_proof_LHS}
    \begin{split}
        &\ddashint_a^b\frac{f(t')}{\sinh^2(\pi T(t'-t))}dt'\\
        &= \frac{1}{\pi T}\frac{d}{dt}\dashint_a^b \coth(\pi T(t'-t))f(t') dt'\\
        &\overset{\eqref{eq:PV_memory_kernel}}{=} \frac{1}{\pi T}\frac{d}{dt}\left(\frac{f(t)}{\pi T} \log\abs{\frac{\sinh(\pi T(b-t))}{\sinh(\pi T(a-t))}} + \int_a^b\frac{R_1(t';t)}{\tanh(\pi T (t'-t))}\right)\\
        &= \frac{f'(t)}{(\pi T)^2}\log\abs{\frac{\sinh(\pi T(b-t))}{\sinh(\pi T(a-t))}} + \frac{f(t)}{\pi T}\left(\coth(\pi T (a-t))-\coth(\pi T (b-t))\right)\\
        &\quad + \frac{1}{\pi T}\int_a^b\frac{\partial_t R_1(t';t)}{\tanh(\pi T (t'-t))} + \int_a^b\frac{R_1(t';t)}{\sinh^2(\pi T (t'-t))}\\
        &= \frac{f'(t)}{(\pi T)^2}\log\abs{\frac{\sinh(\pi T(b-t))}{\sinh(\pi T(a-t))}} + \frac{f(t)}{\pi T}\left(\coth(\pi T (a-t))-\coth(\pi T (b-t))\right)\\
        &\quad - f'(t) \int_a^b\left(\frac{1}{\pi T\tanh(\pi T (t'-t))} - \frac{t'-t}{\sinh^2(\pi T (t'-t))}\right) + \int_a^b\frac{R_2(t';t)}{\sinh^2(\pi T (t'-t))}\\
        &= \frac{f'(t)}{(\pi T)^2}\log\abs{\frac{\sinh(\pi T(b-t))}{\sinh(\pi T(a-t))}} + \frac{f(t)}{\pi T}\left(\coth(\pi T (a-t))-\coth(\pi T (b-t))\right)\\
        &\quad - \frac{f'(t)}{\pi T} \left((t'-t)\coth(\pi T (t'-t))\right)\Big\vert_a^b + \int_a^b\frac{R_2(t';t)}{\sinh^2(\pi T (t'-t))}\\
    \end{split}
\end{equation}
The RHS is
\begin{equation}\label{eq:FP_kernel_proof_RHS}
    \begin{split}
        &\lim_{\varepsilon\to 0}\int_a^b \frac{\left(\sinh^2(\pi T(t'-t))(1-2\varepsilon^2) - \varepsilon^2\right)}{(\sinh^2(\pi T(t'-t)) + \varepsilon^2)^2}f(t') dt'\\
        &= f(t)\lim_{\varepsilon\to 0}\int_a^b \frac{\left(\sinh^2(\pi T(t'-t))(1-2\varepsilon^2) - \varepsilon^2\right)}{(\sinh^2(\pi T(t'-t))+ \varepsilon^2)^2} 
        + f'(t)\lim_{\varepsilon\to 0}\int_a^b \frac{(t'-t)\left(\sinh^2(\pi T(t'-t))(1-2\varepsilon^2) - \varepsilon^2\right)}{(\sinh^2(\pi T(t'-t)) + \varepsilon^2)^2}\\
        &\quad + \int_a^b dt'\frac{R_2(t';t)}{\sinh^2(\pi T(t'-t))}\\
        &= \lim_{\varepsilon\to 0}\Bigg\{f(t) T \left(\frac{\sinh(2\pi T (a-t))}{2\sinh^2(\pi T (a-t)) + 2\varepsilon^2}-\frac{\sinh(2\pi T (b-t))}{2\sinh^2(\pi T (b-t)) + 2\varepsilon^2}\right)\\
        &\quad + f'(t) \bigg[\frac{1}{2\pi} \log\abs{\frac{\sinh^2(\pi T(b-t))+\varepsilon^2}{\sinh^2(\pi T(a-t))+\varepsilon^2}}+T\left(\frac{(a-t)\sinh(2\pi T (a-t))}{2\sinh^2(\pi T (a-t)) + 2\varepsilon^2}-\frac{(b-t)\sinh(2\pi T (b-t))}{2\sinh^2(\pi T (b-t)) + 2\varepsilon^2}\right)\bigg]\Bigg\}\\
        &\quad + \int_a^b dt'\frac{R_2(t';t)}{\sinh^2(\pi T(t'-t))}\\
        &\overset{t\neq a,b}{=}\frac{f(t)}{\pi T}\left(\coth(\pi T (a-t))-\coth(\pi T (b-t))\right) + \frac{f'(t)}{(\pi T)^2}\log\abs{\frac{\sinh(\pi T(b-t))}{\sinh(\pi T(a-t))}}\\
        &\quad  - \frac{f'(t)}{\pi T}\left((b-t)\coth(\pi T (b-t)) - (a-t)\coth(\pi T (a-t))\right) + \int_a^b dt'\frac{R_2(t';t)}{\sinh^2(\pi T(t'-t))}
    \end{split}
\end{equation}
which is identical to \eqref{eq:FP_kernel_proof_LHS}. Notice that for the HFP, we had to take a derivative of $R_1$ requiring $f$ to be differentiable and Hölder continuity is now required for $R_2$.
%\end{widetext}

%% file: sections/appendices/finite_part.tex
The Hadamard finite part (HFP) generalizes the Cauchy principal value (CPV) in the sense that it assigns values to integrals that diverge "worse" than the Cauchy principal value can handle, so-called hypersingular integrals \cite{gelfand_generalized_1993,kutt_numerical_1975,monegato_definitions_2009}. The simplest example for such an integral is $\int dt/t^2$. The symmetry around the singularity prevents the CPV from being able to render it finite. As a rule of thumb, the HFP is obtained by excluding a symmetric region, $(-\varepsilon,\varepsilon)$ around the singularity, computing the expansion of the integral for $\varepsilon\to 0$, and simply discarding the divergent terms \cite{monegato_definitions_2009}. For the example above with $a<0<b$, we isolate the divergences
\begin{equation}\label{eq:app_seperate_divergence}
    \lim_{\varepsilon\to 0} \int_{(a,b)\setminus (-\varepsilon,\varepsilon)}\frac{dt}{t^2} = \lim_{\varepsilon\to 0} \left(\frac{1}{a} + \frac{2}{\varepsilon} - \frac{1}{b}\right)
\end{equation}
and find the HFP, denoted by a twice-dashed integral, by dropping the divergent term,
\begin{equation}\label{eq:app_finite_part_example}
    \ddashint_a^b \frac{dt}{t^2} =\frac{1}{a}-\frac{1}{b}\,.
\end{equation}
One obtains the same result by formally commuting derivatives and the limit of the CPV \cite{ang_hypersingular_2014}, that is
\begin{equation}\label{eq:differentiation_relation}
    \ddashint_a^b \frac{dt}{t^2} =  \ddashint_a^b \partial_x \frac{1}{t-x}\Big\vert_{x=0} dt = \frac{d}{dx} \dashint_a^b \frac{dt}{t-x}\Bigg\vert_{x=0}\,,
\end{equation}
where the once-dashed integral is used for the CPV. The CPV itself can be written as the derivative of a regular integral, i.e.
\begin{equation}
    \dashint_a^b \frac{1}{t-x}dt = \frac{d}{dx} \int_a^b \log\abs{t-x} dt\,.
\end{equation}
In this sense, the CPV itself may be viewed as the first-order finite part, and higher-order finite parts can be obtained iteratively from lower-order ones by successive differentiation, where the zeroth-order finite part is the regular integral.
More rigorously, the HFP of $\int dt f(t)/(t-x)^p$ (for sufficiently smooth functions $f$ \cite{martin_hypersingular_1996}) can be viewed as the meromorphic continuation of the distribution generated by $1/(t-x)^p$ from $\Re p<1$ to $\Re p\geq 1$  \cite{riesz_integrales_1938,gelfand_generalized_1993,blanchet_hadamard_2000}. This formal viewpoint draws a parallel to the more familiar “trick” of analytically continuing the geometric series $\sum_{n=0}^\infty q^n = \frac{1}{1-q}$ to $\abs{q}>1$. In particular, meromorphic continuations of distributions are unique \cite{gelfand_generalized_1993}, and the HFP is therefore not arbitrary like, e.g., the rearrangement of conditionally converging series' partial sums, even though what we did to get from \eqref{eq:app_seperate_divergence} to \eqref{eq:app_finite_part_example} may look suspicious. Finite part integrals inherit linearity from the regular integral. Still, several other intuitive qualities of the standard integral do not hold for them, e.g., $f(x)\geq 0 \Rightarrow \int_a^b f(x) dx \geq 0$ as \eqref{eq:app_finite_part_example} already shows. An overview of these can be found in \cite{kutt_numerical_1975}.\\
More generally we denote by $\fp \left\{K(t'-t)\right\}$ the distribution that maps a test function $h$ to the finite-part of the convolution of $K$ with $h$ over some domain $\mathcal{D}$,
\begin{equation}
    h\mapsto\ddashint_\mathcal{D}K(t'-t)h(t')dt'\,.
\end{equation}
For this to be a well-defined distribution, it is sufficient that the kernel $K$ is a meromorphic function, since then one can use its Laurent series representation and linearity of the integral to recover the monomial HFPs (see “canonical regularization” in \cite{gelfand_generalized_1993} for a more general discussion). For example, in the main text we use
\begin{equation}
    K(t'-t) = \frac{1}{\sinh^2(t'-t)} = \frac{1}{(t'-t)^2} + f(t'-t)\,,
\end{equation} 
where $f$ is analytic. By linearity of the HFP, as a distribution
\begin{equation}
    \fp \frac{1}{\sinh^2(t'-t)} = \fp \frac{1}{(t'-t)^2} + f(t'-t)\,.
\end{equation}
For the numerical implementation of HFPs, the following identity, obtained by “peeling off” the singularity via Taylor expansion, is useful \cite{kutt_numerical_1975,linz_approximate_1985}.
\begin{equation}\label{eq:peel-off}
\begin{split}
    \ddashint_a^b \frac{f(t)}{(t-x)^p} dt &= \sum_{n=0}^{p-1} \frac{f^{(n)}(x)}{n!} \ddashint_a^b \frac{dt}{(t-x)^{p-n}} + \int_a^b  \frac{R_p(t;x)}{(t-x)^p}
\end{split}
\end{equation}
with the remainder $R_p(t) = f(t) - \sum_{n=0}^{p-1} \frac{f^{(n)}(x)}{n!} (t-x)^n$. The last integral is regular if the $(p-1)$-th derivative of $f$ is Hölder continuous around $x$ i.e. $\exists C\geq 0,\alpha>0 :\abs{f^{(p-1)}(t) - f^{(p-1)}(x)}\leq C \abs{t-x}^\alpha$. Then one can use that the remainder
\begin{equation}
    R_p(t;x) = \int_x^t dt' \frac{(t-t')^{p-2}}{(p-2)!}\left(f^{(p-1)}(t')-f^{(p-1)}(x)\right)
\end{equation}
can be bounded using the triangle inequality and the Hölder condition as
\begin{equation}
    \abs{R_p(t;x)} \leq C \int_x^t dt' \frac{(t-t')^{p-2}\abs{t'-x}^\alpha}{(p-2)!}
\end{equation}
and therefore
\begin{equation}
     \frac{R_p(t;x)}{(t-x)^p} \sim \mathcal{O}\left((t-x)^{\alpha-1}\right)
\end{equation}
is integrable around $x$. The remainder term in \eqref{eq:peel-off} is a regular integral that can be implemented numerically, while the finite number of HFP integrals multiplying the first $p$ Taylor coefficients can be solved analytically.\\
Like the CPV, the HFP can be written as a limit of distributions generated by a mollifier. For example, differentiating the well-known (e.g., from the Sokhotski–Plemelj theorem) nascent CPV of $1/t$,
\begin{equation}\label{eq:nascent_CPV}
    \dashint_a^b \frac{1}{t'-t} dt' = \lim_{\varepsilon\to 0} \int_a^b \frac{t'-t}{(t'-t)^2 + \varepsilon^2}dt'
\end{equation}
we find using \eqref{eq:differentiation_relation} and \eqref{eq:nascent_CPV}
\begin{equation}\label{eq:nascent_HFP}
    \ddashint_a^b \frac{1}{(t'-t)^2}dt' = \lim_{\varepsilon\to 0} \int_a^b \frac{(t'-t)^2-\varepsilon^2}{((t'-t)^2+\varepsilon^2)^2}dt'
\end{equation}
In the main text, such integrands are referred to as nascent HFPs. In the next section, we will show that the Fourier transform to real space of the regulated symmetric embedding self-energy is precisely such a nascent HFP.\\
Lastly, some authors also define “one-sided” finite parts, i.e., finite part integrals where the singularity lies on the boundary of the domain of integration\cite{kutt_numerical_1975, gelfand_generalized_1993,monegato_definitions_2009}. These are defined analogously by meromorphic continuation. However, these distributions are fundamentally different objects. In particular, a nascent "two-sided" HFP, e.g. \eqref{eq:nascent_HFP} does not converge to the one-sided HFP if it is used on a domain with the singularity on its boundary. In fact, it does not converge at all, as can be seen by evaluating
\begin{equation}
    \lim_{\varepsilon\to 0} \int_a^b \frac{(t'-t)^2-\varepsilon^2}{((t'-t)^2+\varepsilon^2)^2} =\lim_{\varepsilon\to 0} \left[\frac{t-t'}{(t'-t)^2 + \varepsilon^2} \right]^b_a 
\end{equation}
and setting either $a=t$ or $b=t$.

%% file: sections/appendices/nascent_delta2.tex
Given any absolutely integrable function $K\in L_1(\mathbb{R})$ with $\int_\mathbb{R} K = 1$ (by theorem 9.8 in \cite{wheeden_measure_1977}),
\begin{equation}
    K_\varepsilon(x) := \frac{1}{\varepsilon}K\left(\frac{x}{\varepsilon}\right) \overset{\varepsilon\rightarrow 0^+}{\longrightarrow} \delta(x) \quad \text{as a distribution.}
\end{equation}
In particular, the function $x\mapsto\frac{1}{(x^2+1)^2}$ is absolutely integrable and integrates to
\begin{equation}
    \int_{-\infty}^\infty\frac{1}{(x^2+1)^2} = \left[\frac{x}{2(x^2+1)}+\frac{1}{2}\arctan(x)\right]_{-\infty}^\infty = \frac{\pi}{2}
\end{equation}
such that
\begin{equation}
    K_\epsilon(x) = \frac{2}{\pi}\frac{\varepsilon^3}{(x^2+\varepsilon^2)^2}\rightarrow \delta(x)
\end{equation}
and therefore
\begin{equation}
    \frac{\varepsilon^2}{(x^2+\varepsilon^2)^2} \overset{\varepsilon\to 0}{\longrightarrow}\frac{\pi}{2\varepsilon}\delta(x)
\end{equation}

%% file: sections/appendices/zero_and_one_terms.tex
\onecolumngrid
In this appendix, we state the analytic integral solutions for the zero-order term, $P(t_1,t_2)$, and the first-order term, $Q(t_1,t_2)$, which appear in \eqref{eq:memory_term_interpretation}. Furthermore, the limits for times far from the initial time are shown, in which case both terms become translation invariant functions, i.e., functions of $t_1 - t_2$. The zero-order term is given by
%\small
\begin{equation}
\begin{split}
    P(t_1,t_2) &= \int_{t_0}^{t_2} \Sigma_\emb^S(t_1,t') dt' = i\gamma T \left(\frac{\sinh(2\pi T (t_2-t_1))}{\cos(\frac{2\pi T}{\omega_c}) - \cosh(2\pi T (t_2-t_1))}-\frac{\sinh(2\pi T (t_0-t_1))}{\cos(\frac{2\pi T}{\omega_c}) - \cosh(2\pi T (t_0-t_1))}\right)\\
    &\quad \xrightarrow{\abs{t_0-t_1} \gg \omega_c\inv} i\gamma T \left(\frac{\sinh(2\pi T (t_2-t_1))}{\cos(\frac{2\pi T}{\omega_c}) - \cosh(2\pi T (t_2-t_1))} + \coth(\pi T (t_0-t_1)) \right)\\
    &\quad \xrightarrow{\pi T (t_0-t_1)\ll -1} i\gamma T \left(\frac{\sinh(2\pi T (t_2-t_1))}{\cos(\frac{2\pi T}{\omega_c}) - \cosh(2\pi T (t_2-t_1))} - 1 \right)
\end{split}
\end{equation}%\normalsize
The first-order term is given by
%\small
\begin{equation}
    \begin{split}
        Q(t_1,t_2) &= \int_{t_0}^{t_2} \Sigma_ \emb^S(t_1,t')(t'-t_1) dt'\\
        &= \frac{i\gamma}{2\pi} \log(\frac{\cos(\frac{2\pi T}{\omega_c}) - \cosh(2\pi T (t_2-t_1))}{\cos(\frac{2\pi T}{\omega_c}) - \cosh(2\pi T (t_0-t_1))})\\
        &\quad + i\gamma T \left(\frac{(t_2-t_1)\sinh(2\pi T (t_2-t_1))}{\cos(\frac{2\pi T}{\omega_c}) - \cosh(2\pi T (t_2-t_1))}-\frac{(t_0-t_1)\sinh(2\pi T (t_0-t_1))}{\cos(\frac{2\pi T}{\omega_c}) - \cosh(2\pi T (t_0-t_1))}\right)\\
        &\quad \overset{2\pi T (t_0-t_1)\ll -1}{\longrightarrow} \frac{i\gamma}{2\pi} \log(2\cosh(2\pi T (t_2-t_1))-2\cos(\frac{2\pi T}{\omega_c})) + i\gamma T \frac{(t_2-t_1)\sinh(2\pi T (t_2-t_1))}{\cos(\frac{2\pi T}{\omega_c}) - \cosh(2\pi T (t_2-t_1))}
    \end{split}
\end{equation}
where we used that $\left[2x\coth{x}-\log(2\sinh(x)^2)\right] \to \log 2$ as $x\to -\infty$, for the limit.\\
Unfortunately, while readable, the expressions above are not suited for numerical implementation as they suffer from overflow. Let us therefore define $t_{ij}:=2\pi T_B (t_i-t_j)$, $z_{ij} := e^{-\abs{t_{ij}}}$ and $\Delta := \cos(\frac{2\pi T}{\omega_c})$. Then we find, using
\begin{equation}
    \cosh(t)=\cosh(\abs{t}) = \frac{1+z^2}{2z}\quad\text{and}\quad \sinh(t) = \sgn(t) \sinh(\abs{t}) = \sgn(t)\frac{1-z^2}{2z}
\end{equation} in the same limit as above
\begin{equation}
\begin{split}
    P(t_1,t_2) &= i\gamma T_B\left(\sgn(t_{21})\frac{1-z_{21}^2}{2\Delta z_{21} - z_{21}^2 -1}-\sgn(t_{01})\frac{1-z_{01}^2}{2\Delta z_{01} - z_{01}^2 -1}\right)\\
    &\rightarrow i\gamma T_B\left(\sgn(t)\frac{1-z_{21}^2}{2\Delta z_{21} - z_{21}^2 -1}-1\right)
\end{split}
\end{equation}
and
\begin{equation}
\begin{split}
    Q(t_1,t_2) &= i\frac{\gamma}{2\pi}\left(\abs{t_{21}}\frac{1-z_{21}^2}{2\Delta z_{21} - z_{21}^2 -1} - \abs{t_{01}}\frac{1-z_{01}^2}{2\Delta z_{01} - z_{01}^2 -1} + \abs{t_{21}}-\abs{t_{01}} + \log(\frac{2\Delta z_{21}-1-z_{21}^2  }{2\Delta z_{01}-1-z_{01}^2})\right)\\
    &\to i\frac{\gamma}{2\pi}\left(\abs{t_{21}}\frac{1-z_{21}^2}{2\Delta z_{21} - z_{21}^2 -1} + \abs{t_{21}}+ \log(1+z_{21}^2 - 2\Delta z_{21})\right)
\end{split}
\end{equation}
such that no exponentially growing terms appear. In the same spirit, we also implement the memory kernel as
\begin{equation}
    \Sigma^S(t_1,t_2) = i\gamma 2\pi T_B^2 \frac{\Delta - \sech(t)}{\cosh(t)(\Delta \sech(t) - 1)^2}
\end{equation}
\begin{figure}[H]
    \centering
    \includegraphics[width=0.95\linewidth]{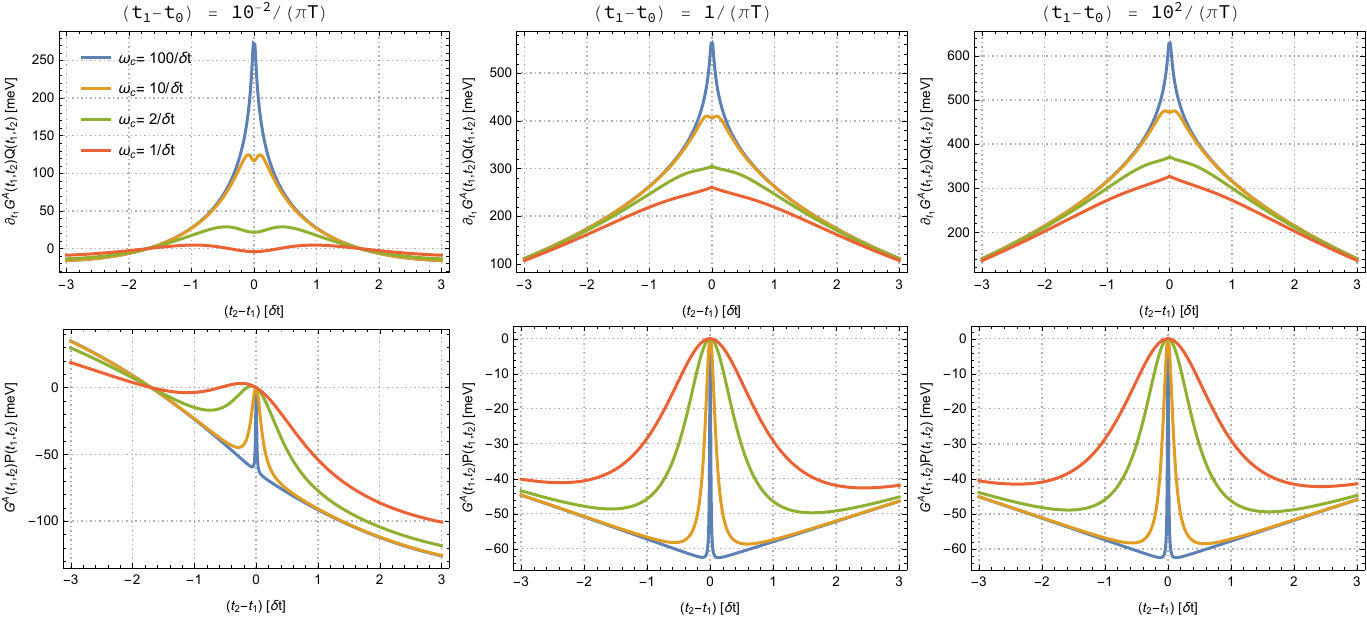}
    \caption{The zeroth- and first-order terms in the memory kernel expansion are shown at different distances from the initial times (the three columns) around the diagonal for $\gamma = 200$ meV, $\omega_\gamma = 200$ meV, $T=1$ meV, and $\delta t = 2\pi/(30\, \omega_\gamma)$. Additionally, for each time, the function is shown with the bath responsiveness ranging from the scale of the system time-discretization, to two orders of magnitude above that. It can be seen that with increasing distance from the initial time, both functions become time translation invariant i.e. functions of the relative time $t_1-t_2$. Furthermore, for $\omega_c \gtrsim 10/\delta t$, the function values are indistinguishable from those of the $\omega_c\to\infty$ limit for all off-diagonal nodes.}
    \label{fig:Zero_and_One_overview}
\end{figure}
\normalsize

%% file: sections/appendices/ETD.tex
\onecolumngrid
Let us define $u^T =  \begin{pmatrix}G^S & \partial_{t_1}G^S\end{pmatrix}$, then we can write the symmetric KBE in the limit $\omega_c \to \infty$, where the dissipation term becomes local, as
\begin{equation}
    \partial_{t_1}\begin{pmatrix}G^S \\ \partial_{t_1}G^S\end{pmatrix} = \begin{pmatrix}
        0 & 1 \\
        -\omega_0^2 & -\gamma
    \end{pmatrix}\begin{pmatrix}G^S \\ \partial_{t_1}G^S\end{pmatrix} + \begin{pmatrix}
        0 \\
        -\Sigma^S \cdot G^\adv - \Sigma^\urcorner\star G^\ulcorner
    \end{pmatrix} \quad\text{in short}\quad  u' = -\text{A} u + f
\end{equation}
A time step is then provided by the exponential quadrature method \cite{hochbruck_exponential_2010}
\begin{equation}
    u_{n+1} = e^{-\text{A}\Delta t} u_n + \int_0^{\Delta t} e^{-(\Delta t - \tau) \text{A}} f(t_n + \tau) d\tau
\end{equation}
The matrix exponential is easily computed to yield
\begin{equation}
    \exp(-\text{A} t) = \begin{pmatrix}
        e^{-\frac{\gamma}{2} t} \left(\cos(\omega_\gamma t) + \frac{\gamma}{2\omega_\gamma}\sin(\omega_\gamma t)\right) & e^{-\frac{\gamma}{2} t} \frac{\sin(\omega_\gamma t)}{\omega_\gamma}\\
         - e^{-\frac{\gamma}{2} t} \frac{\omega_0^2}{\omega_\gamma}\sin(\omega_\gamma t) & e^{-\frac{\gamma}{2} t} \left(\cos(\omega_\gamma t) - \frac{\gamma}{2\omega_\gamma}\sin(\omega_\gamma t)\right)
    \end{pmatrix}
\end{equation}
Stepping from $u_n$ to both $u_{n+1}$ and $u_{n-1}$, one finds that the "velocity" dependence cancels upon adding $u_{n+1} + e^{-\gamma \Delta t} u_{n-1}$ and we can rearrange the result to the "position-only" ETD step
\begin{equation}
\begin{split}
    G^S(t_{n+1},t_m) &= 2 e^{-\frac{\gamma}{2}\Delta t} \cos(\omega_\gamma \Delta t) G^S(t_n,t_m) - e^{-\gamma \Delta t} G^S(t_{n-1},t_m)\\
    &\quad + \int_{-\Delta t}^{\Delta t} e^{-\frac{\gamma}{2}(\Delta t - \tau )} \frac{1}{\omega_\gamma}\sin\left(\omega_\gamma (\Delta t - \abs{\tau})\right) \left(-\Sigma^S \cdot G^\adv - \Sigma^\urcorner\star G^\ulcorner\right)(t_n + \tau, t_m)d\tau
\end{split}
\end{equation}
The USP of this scheme is that the linear part, containing also the damping, is solved exactly. If there was no inhomogeneity the solution would be exact to machine precision. This is not surprising since that equation is analytically solvable and the ETD exploits this. The method is intended to treat problems where the linear term, A, introduces stiffness into the equation. In our case the stiffness resided in the inhomogeneity, so it looks like we have gained nothing. However, we were able to avoid using a finite difference scheme for the damping term. If we had used a finite difference approximation, even a very high order one, we could have recovered its Whittaker–Shannon interpolation, at best. This would have introduced a cutoff in $\omega$-space at the Nyquist frequency, $\omega_N = \pi/\Delta t$. To ensure equilibration to the correct temperature we must retain the FDR. Since $\omega_c\gg \Delta t\inv$, the fluctuation and dissipation spectra are not band-limited below $\omega_N$, and we would introduce aliasing, i.e. spectral weight being down-folded from frequencies above $\omega_N$ to frequencies below $\omega_N$. The consequence would be effective damping rates and temperatures that differ from the true values of the environment.

%% file: sections/appendices/t0_shifting.tex
In this appendix, we explain the claim that
\begin{equation}\label{eq:claim}
    -i\int_0^{\beta}d\tau \Sigma^\urcorner(t_1,\tau) G^\ulcorner(\tau,t_2) = \int_{-\infty}^{t_0} dt'\left[\Sigma^A(t_1,t')G^S(t',t_2) - \Sigma^S(t_1,t') G^A(t',t_2)\right]
\end{equation}
For this to hold, we must assume that there is some relaxation or finite memory such that 
\begin{equation}
    \lim_{\abs{t_1-t_2}\to\infty} \Sigma(t_1,t_2) \to 0
\end{equation}
in particular also $\Sigma^\urcorner(t,\tau) = \Sigma^<(t,t_0-i\tau) \overset{\abs{t-t_0}\to\infty}{\longrightarrow}0$. Then, for $t'_0<t_0$, we may construct a new contour 
\begin{equation}
    \mathcal{C}' = [t'_0,t_0]\oplus(t_0,\infty)\oplus(\infty,t_0)\oplus[t_0,t'_0)\oplus(t'_0,t'_0-i\beta)
\end{equation}
where the order of appearance indicated the contour ordering from earliest (left) to latest (right), though the vertical track could also go first.
On the original contour, 
\begin{equation}
    \mathcal{C} = [t_0,\infty)\oplus(\infty,t_0)\oplus(t_0,t_0-i\beta)\,,
\end{equation}
the system Hamiltonian $\hat H_s(z)$ is defined as
\begin{equation}
    \hat H_s(z) = \begin{cases}
        \hat H_s(t_\pm)\,,\quad z=t_\pm \in [t_0,\infty)\oplus(\infty,t_0)\\
        \hat H_s^M, \qquad z\in (t_0,t_0-i\beta)
    \end{cases}
\end{equation}
with the total Hamiltonian given by 
\begin{equation}
    \hat H(z) = \hat H_s(z) + \hat H_\text{int} + \hat H_e
\end{equation}
such that the system and environment are initially thermalized as $\rho_\text{eq} = \frac{e^{-\beta(H^M_s + H_\text{int}+H_e)}}{\tr e^{-\beta(H^M_s + H_\text{int}+H_e)}}$, though we are interested mostly in the reduced density matrix, $\tr_e\rho_\text{eq}$ of the system.
On the new contour, we define
\begin{equation}
    \hat H'_s(z) = \begin{cases}
        \hat H_s(t_\pm)\,,\quad z=t_\pm \in [t_0,\infty)\oplus(\infty,t_0)\\
        \hat H_s^M, \qquad z\in (t'_0,t'_0-i\beta)\oplus[t'_0,t_0]\oplus[t_0,t'_0)
    \end{cases}
\end{equation}
with the total Hamiltonian defined as before.
The initial states at $t_0$ and $t'_0$ are identical $\rho(t'_0) = \rho_\text{eq} = \rho(t_0)$ but so are all states between $t'_0$ and $t_0$, i.e.
\begin{equation}
    \hat U(t,t'_0)\hat \rho(t'_0) \hat U(t'_0,t) = \rho(t_0)\quad \text{for}\quad  t'_0 \leq t\leq t_0
\end{equation}
because the initial state is by construction a stationary state under the Hamiltonian governing that time interval (both $\hat \rho$ and $\hat U$ are exponentials of the same operator). Then for $t_1,t_2 \geq t_0$
\begin{equation}\label{eq:t0_shift}
\begin{split}
    iG(t_1,t_2) &= \int D\varphi\, \varphi(t_1) \varphi(t_2) e^{i\int_\mathcal{C}dz L(z)} = \tr{\hat{\rho}(t_0)\hat\varphi_H(t_1)\hat\varphi_H(t_2)}\\
    &= \tr{\hat{\rho}(t'_0)\hat U(t'_0,t_0)\hat\varphi_H(t_1)\hat\varphi_H(t_2)\hat U(t_0,t'_0)} = \int D\varphi \varphi(t_1) \varphi(t_2) e^{i\int_\mathcal{C'}dz L'(z)}
\end{split} 
\end{equation}
Consequently, the KBEs on $\mathcal{C}$ and $\mathcal{C}'$ have the same solution $G(t_1,t_2)$, or rather the solution on $\mathcal{C}'$ restricted to $t_1,t_2\geq t_0$ agrees with the one on $\mathcal{C}$. Then using the assumption of decaying memory, we conclude that
\begin{equation}
\begin{split}
    G_0^{-1} G^S(t_1,t_2)  &\overset{\mathcal{C}'}{=} \int_{t'_0}^{\infty} dt'\left[\Sigma^A(t_1,t')G^S(t',t_2) - \Sigma^S(t_1,t') G^A(t',t_2)\right] -i\int_0^{\beta}d\tau \Sigma^\urcorner(t_1,\tau) G^\ulcorner(\tau,t_2)\\
    &\overset{t'_0\to -\infty}{\longrightarrow} \int_{-\infty}^{\infty} dt'\left[\Sigma^A(t_1,t')G^S(t',t_2) - \Sigma^S(t_1,t') G^A(t',t_2)\right]\,,
\end{split}    
\end{equation}
where we shifted $t_0'$ into the infinite past, while at the same time, it still holds that
\begin{equation}
    G_0^{-1} G^S(t_1,t_2) \overset{\mathcal{C}}{=} \int_{t_0}^{\infty} dt'\left[\Sigma^A(t_1,t')G^S(t',t_2) - \Sigma^S(t_1,t') G^A(t',t_2)\right] -i\int_0^{\beta}d\tau \Sigma^\urcorner(t_1,\tau) G^\ulcorner(\tau,t_2)\,.
\end{equation}
A comparison of terms implies \cref{eq:claim}.
Conceptually, the idea is that, under certain conditions, it makes no difference for the future dynamics how long the system was in the initial state beforehand. For Markovian dynamics, this statement is trivial. For non-Markovian dynamics, this holds for times much longer than the memory scale of the environment. The initial correlations captured by $\Sigma^\urcorner\star G^\ulcorner$ already encode in them a would-be-memory of that state. As shown in \eqref{eq:t0_shift} (and the construction leading op to it), this holds in general, but is only useful if the memory-term and therefore the initial correlations also decay such that the term, $\Sigma^\urcorner\star G^\ulcorner$ in the KBE vanishes for the times of interest $t_1,t_2\geq t_0$. Then one has effectively exchanged one term for another instead of just adding one. As we have seen in the main text, the self-energy $\Sigma^\urcorner$ and the propagator $G^\ulcorner$ do not generally decay over the same time-scale. However, the time that it takes for the contribution $\Sigma^\urcorner\star G^\ulcorner$ to decay is given by the faster of the two scales.\\

% There are not many circumstances in which it is advantageous to go from a vertical contour to one that extends $\sim \tau_{mem}$ into the past, given the step number scaling behavior of time-stepping algorithms. However, in handling the KBE integration on the problematic diagonal $t_1=t_2$, we integrate the expressions $Q$ and $P$ using adaptive quadrature. They are only translationally invariant for times far from the initial time. By exploiting the fact that we can absorb their contribution to $\Sigma^\urcorner\star G^\ulcorner$ into the memory-kernel, we can eliminate their $t_0$ dependence, rendering them translationally invariant. Consequently, we do not need to recalculate the weights for the Filon-type quadrature each step, but we can calculate them once beforehand at negligible cost. For short simulations, the adaptive quadrature integrations are the major bottleneck of the algorithm, such that this trick can result in a $10-100\times$ speed up. 